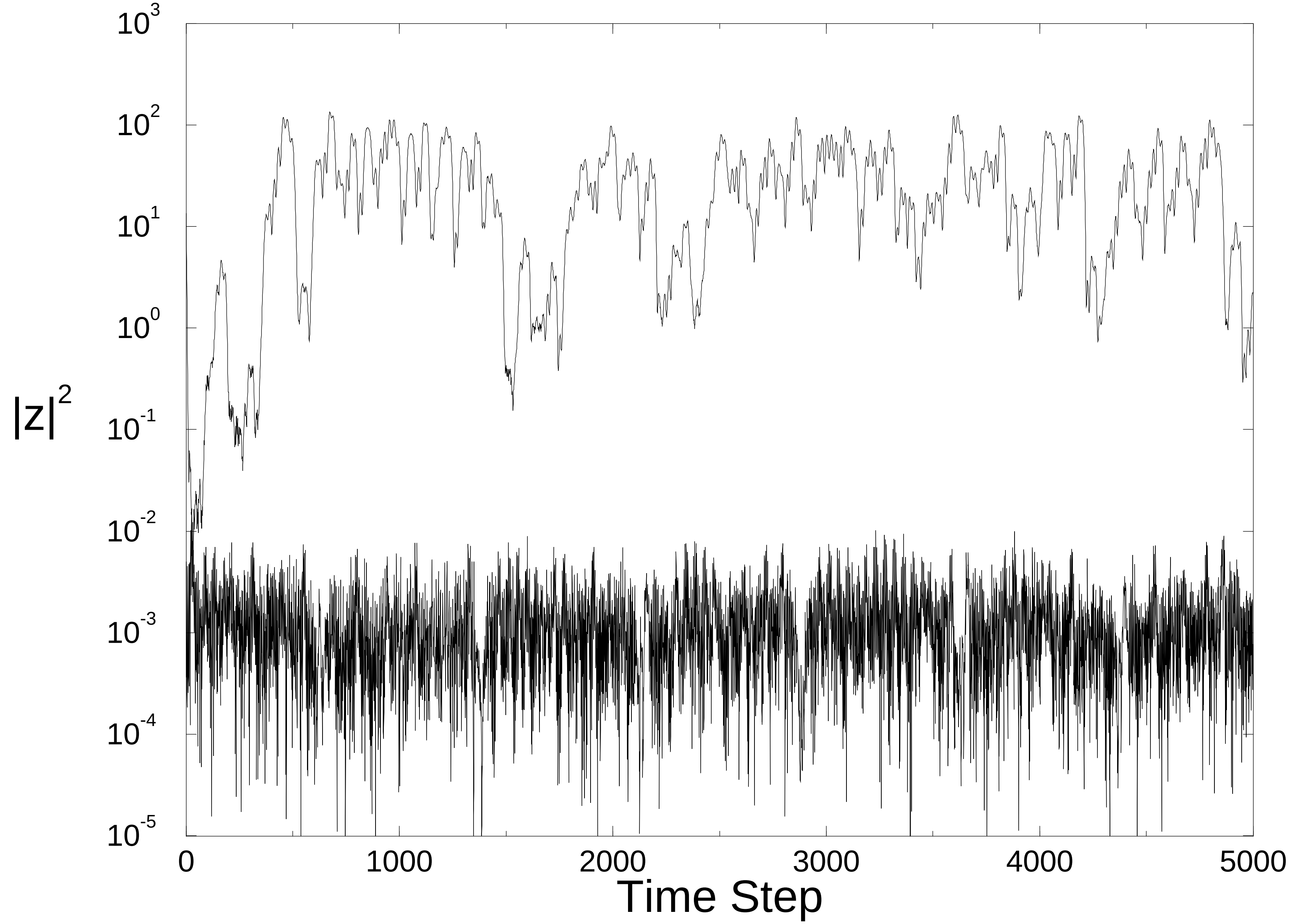

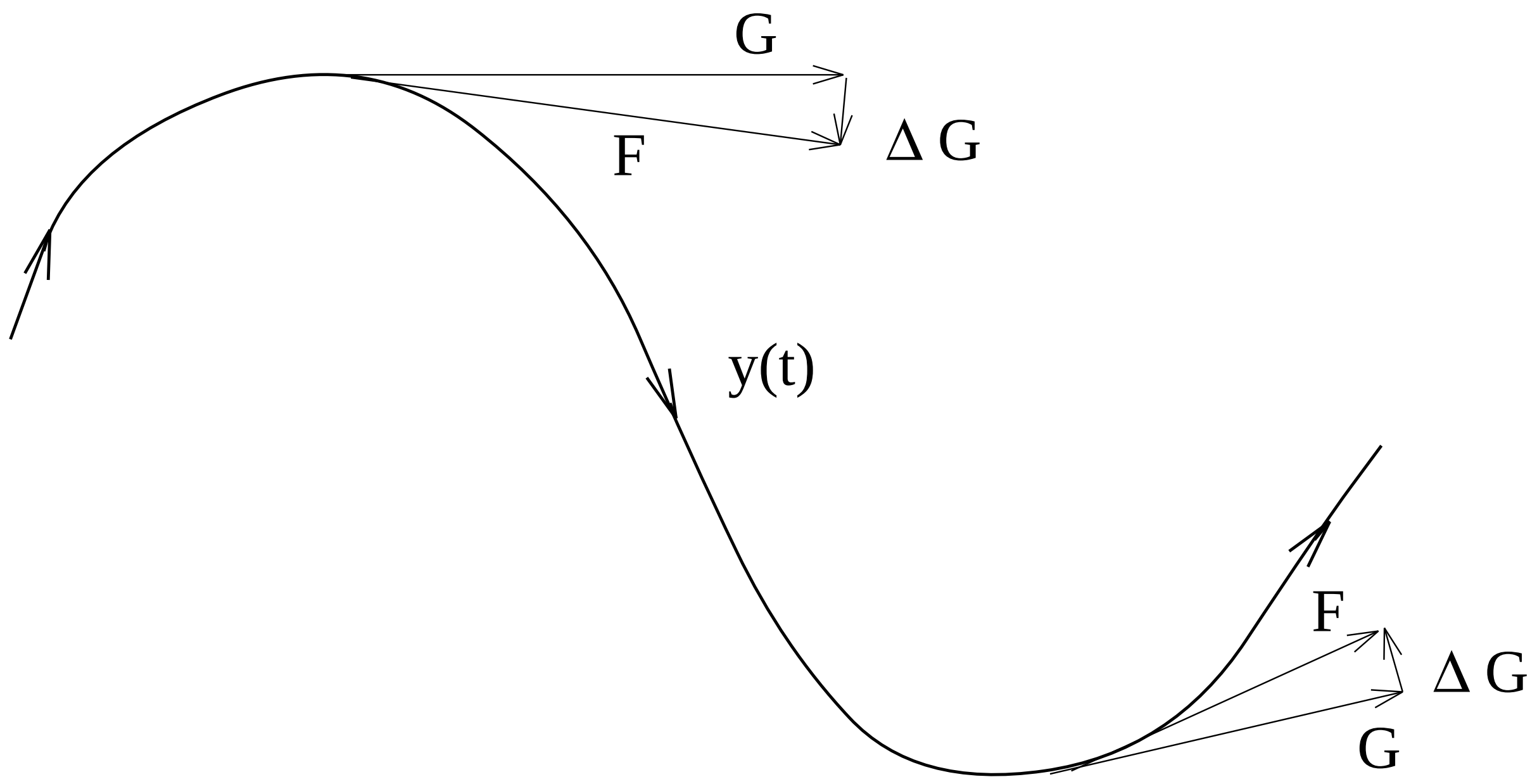

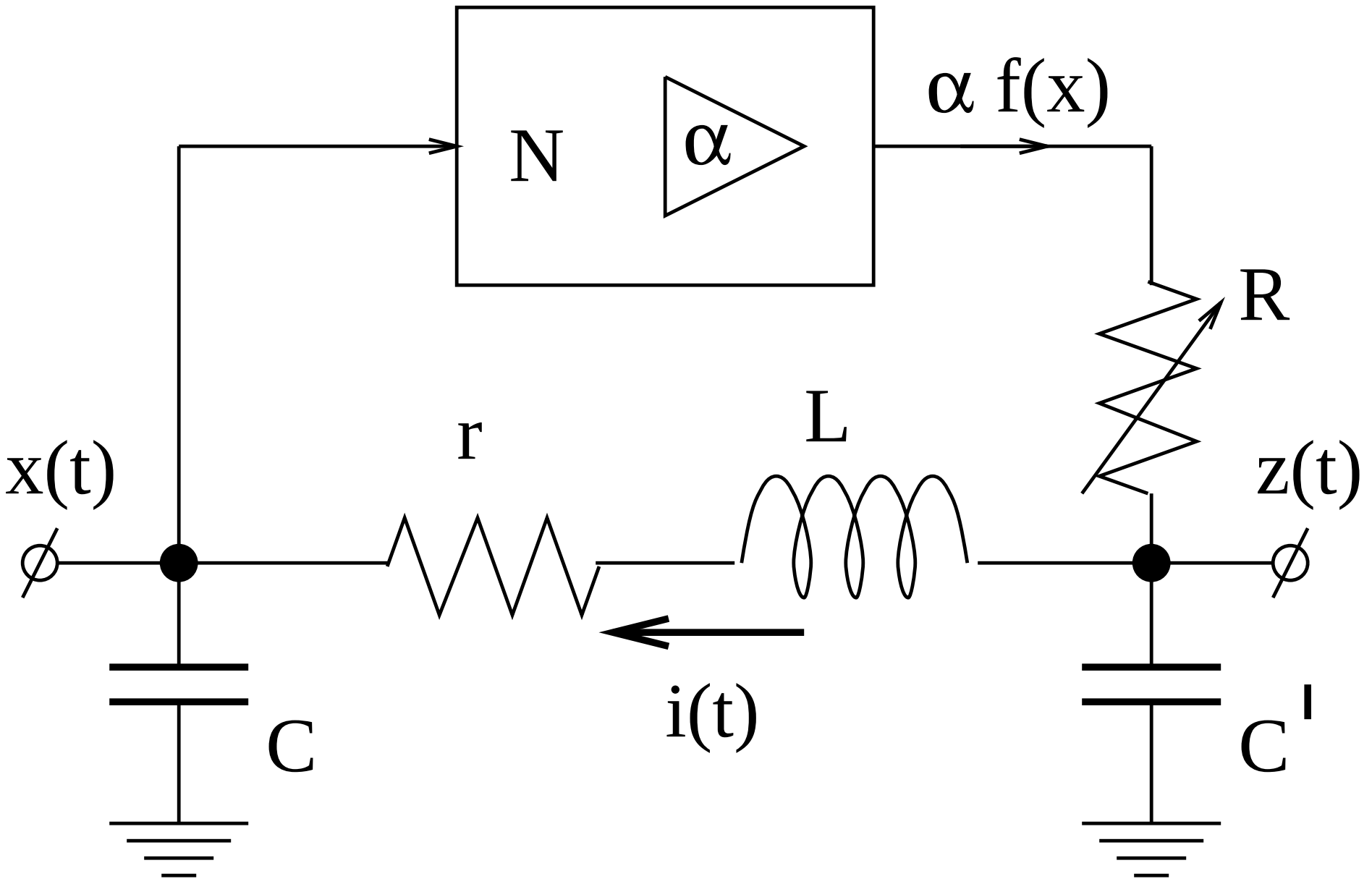

(a)

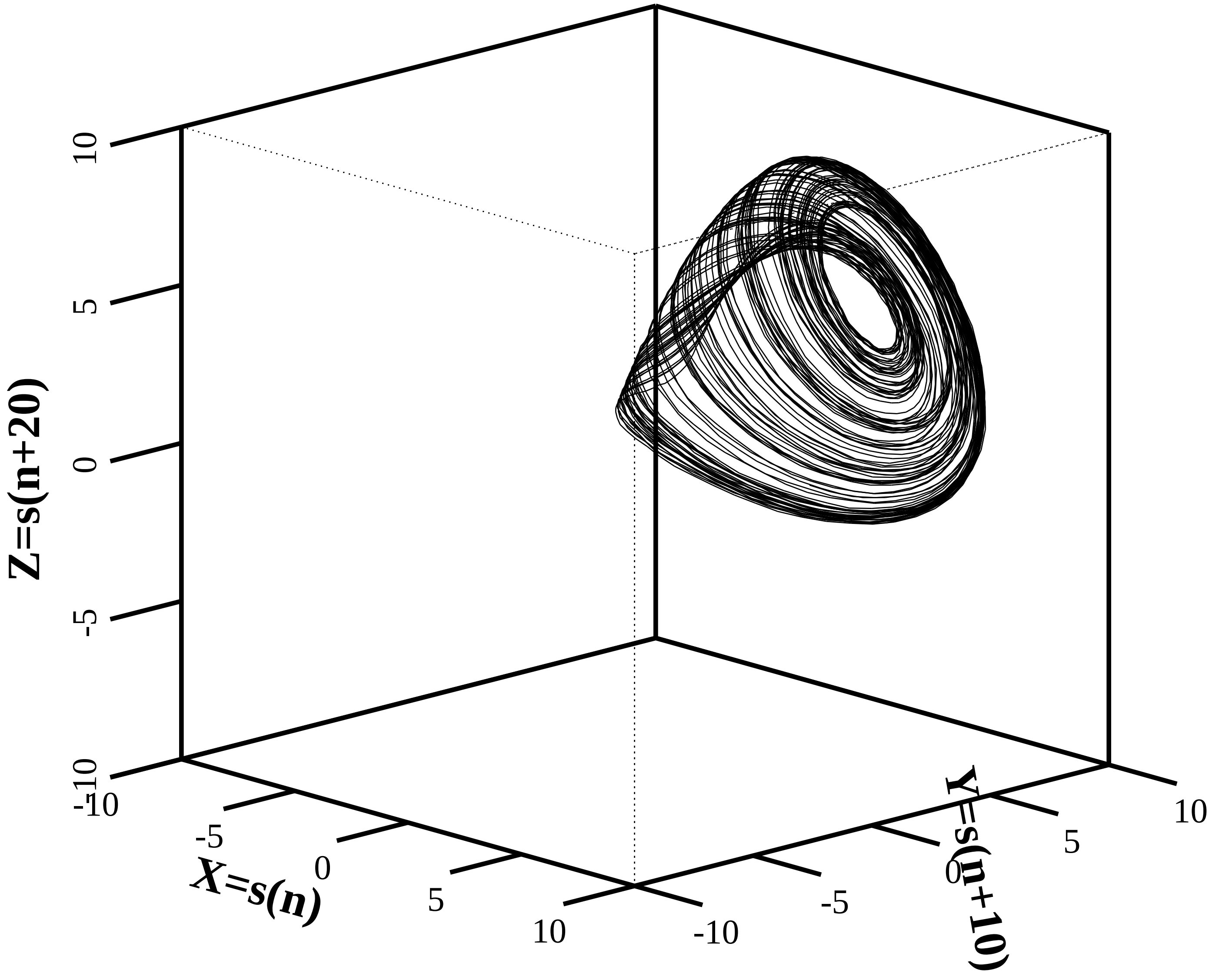

(b)

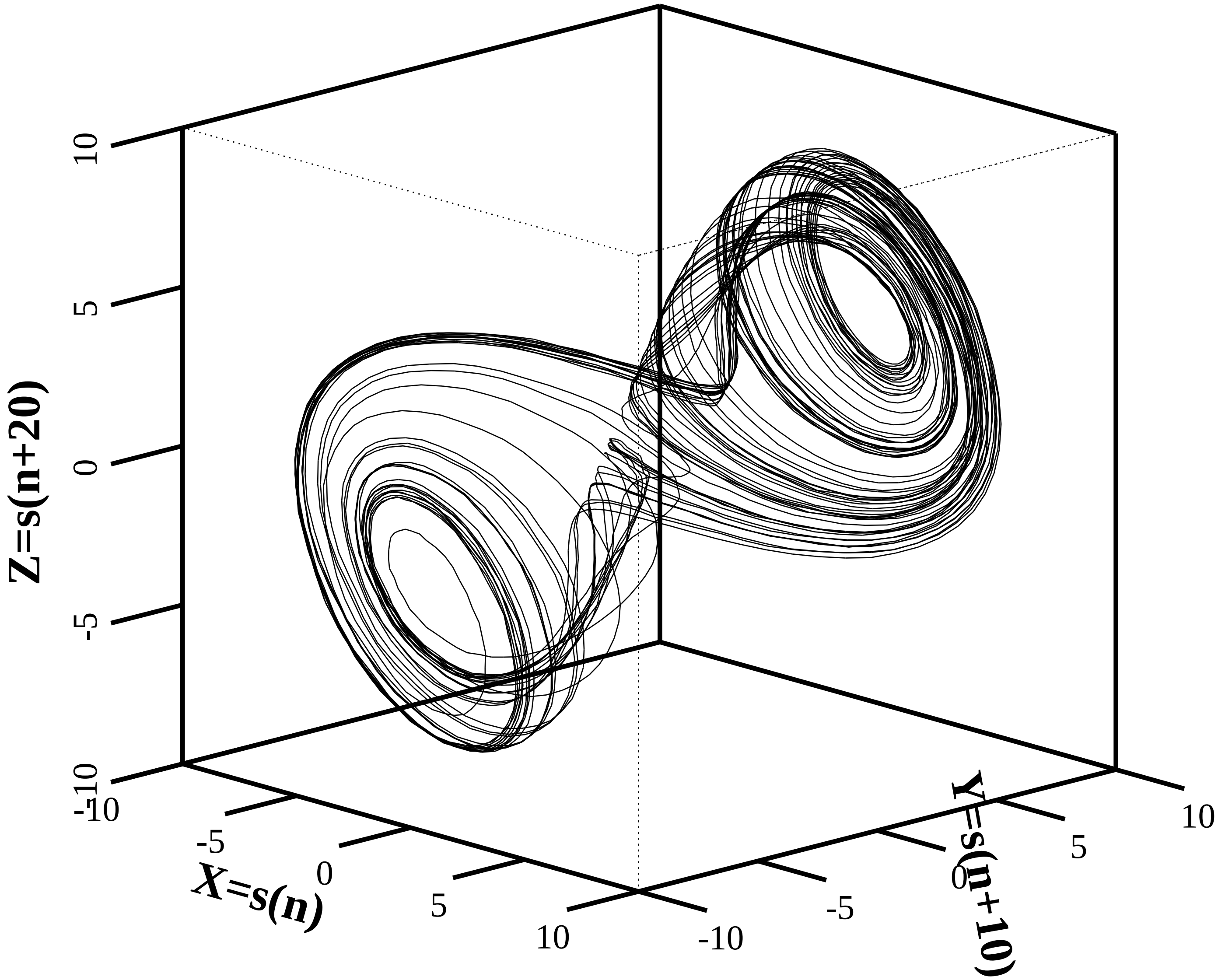

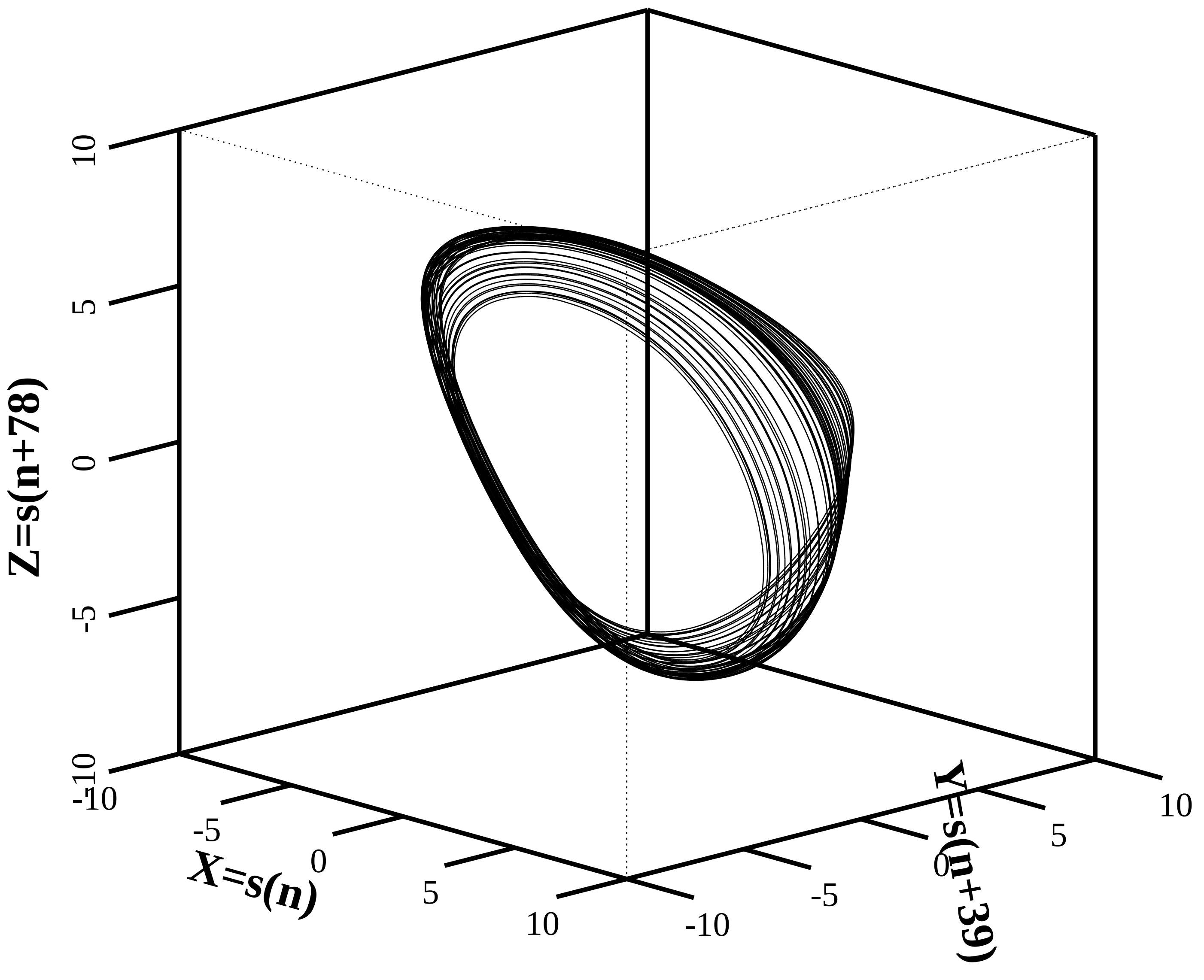

(a)

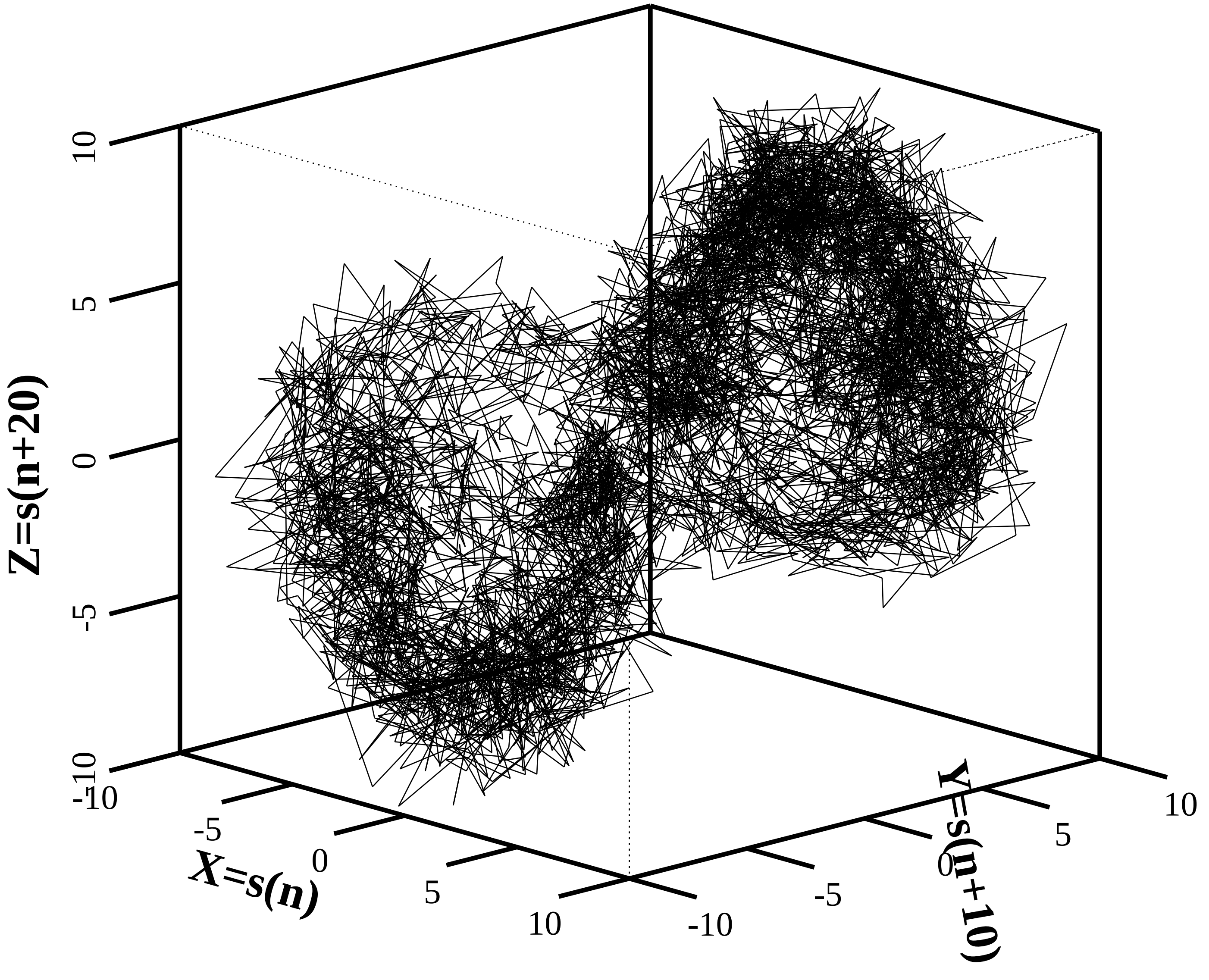

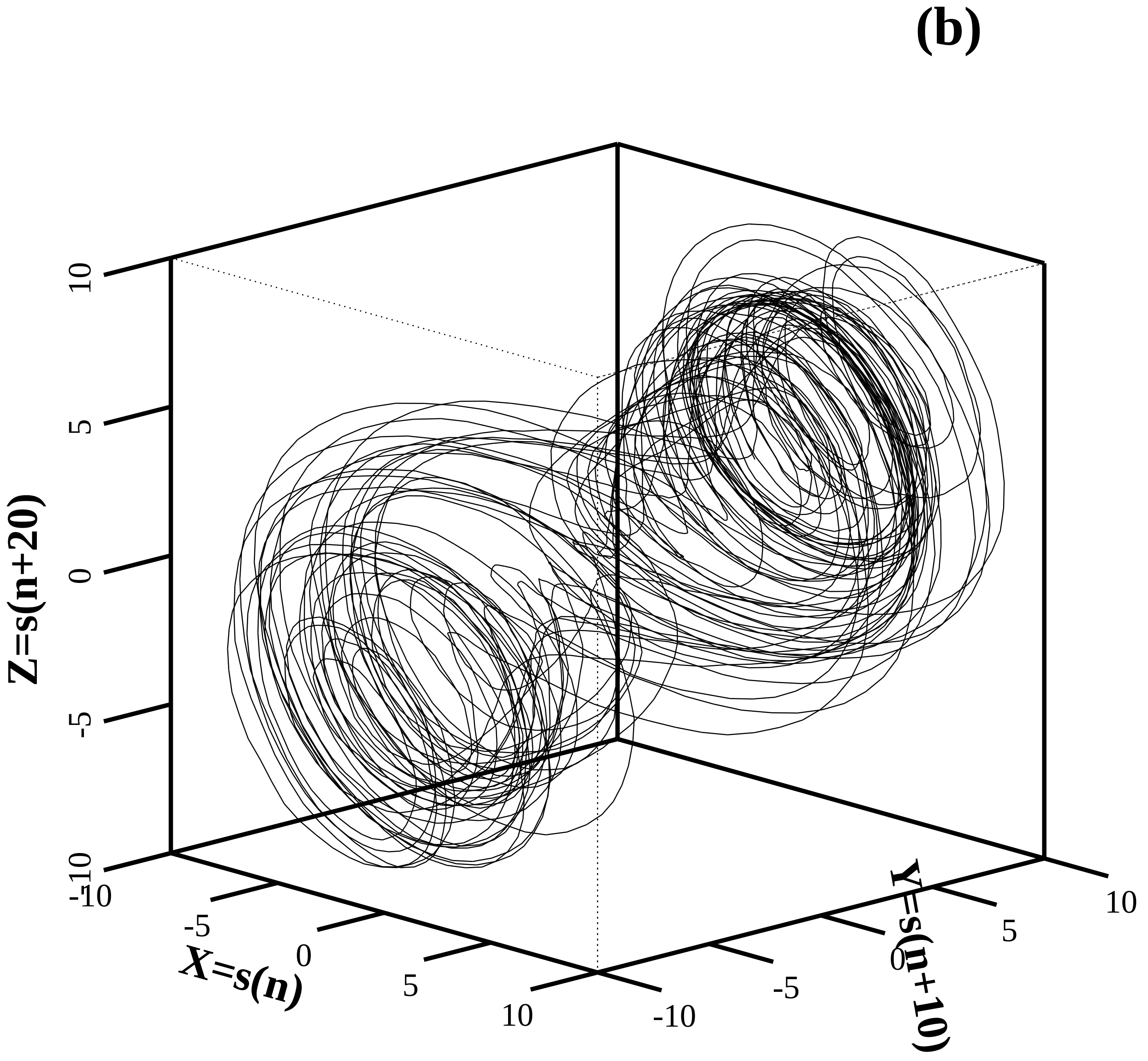

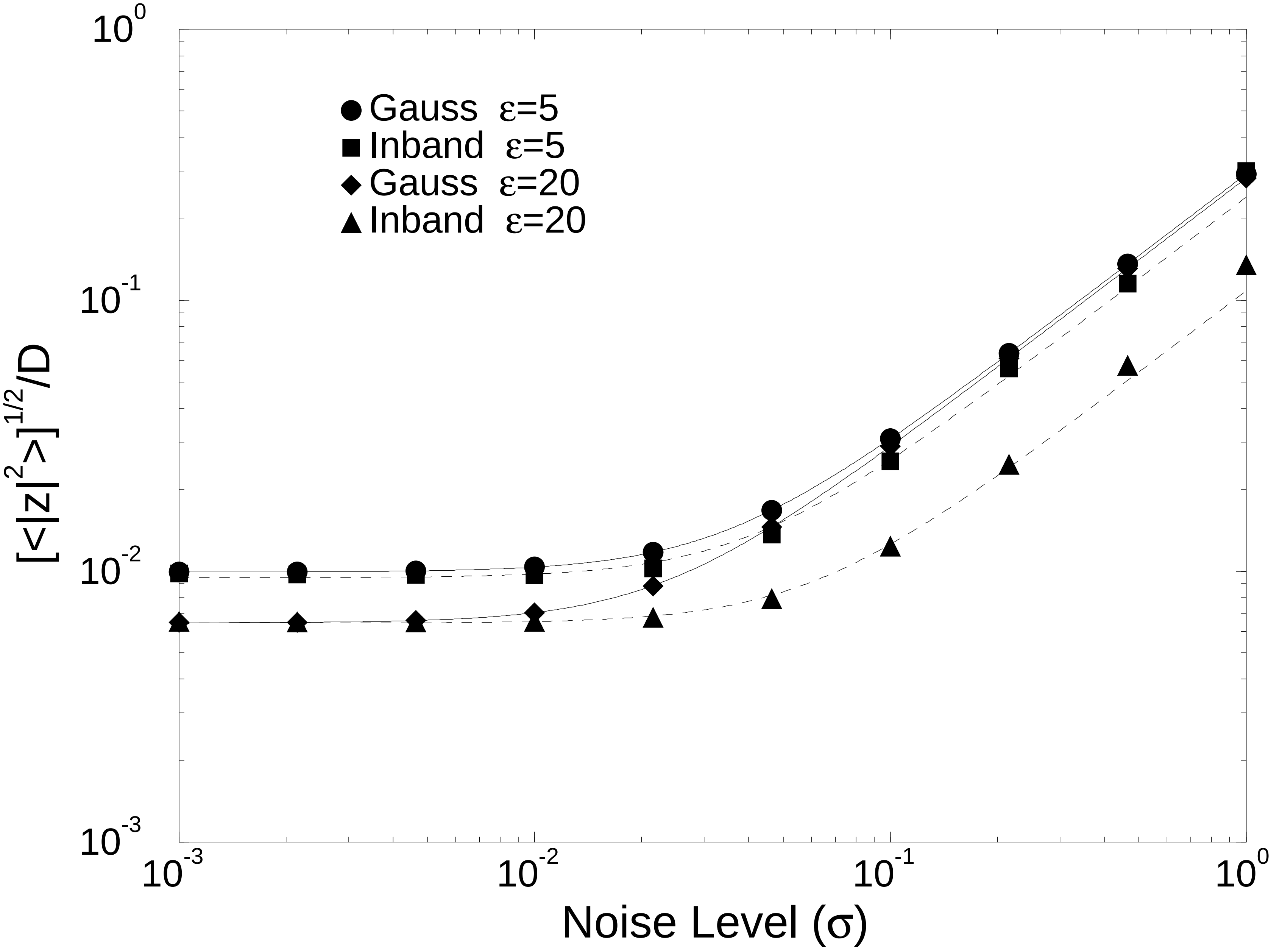

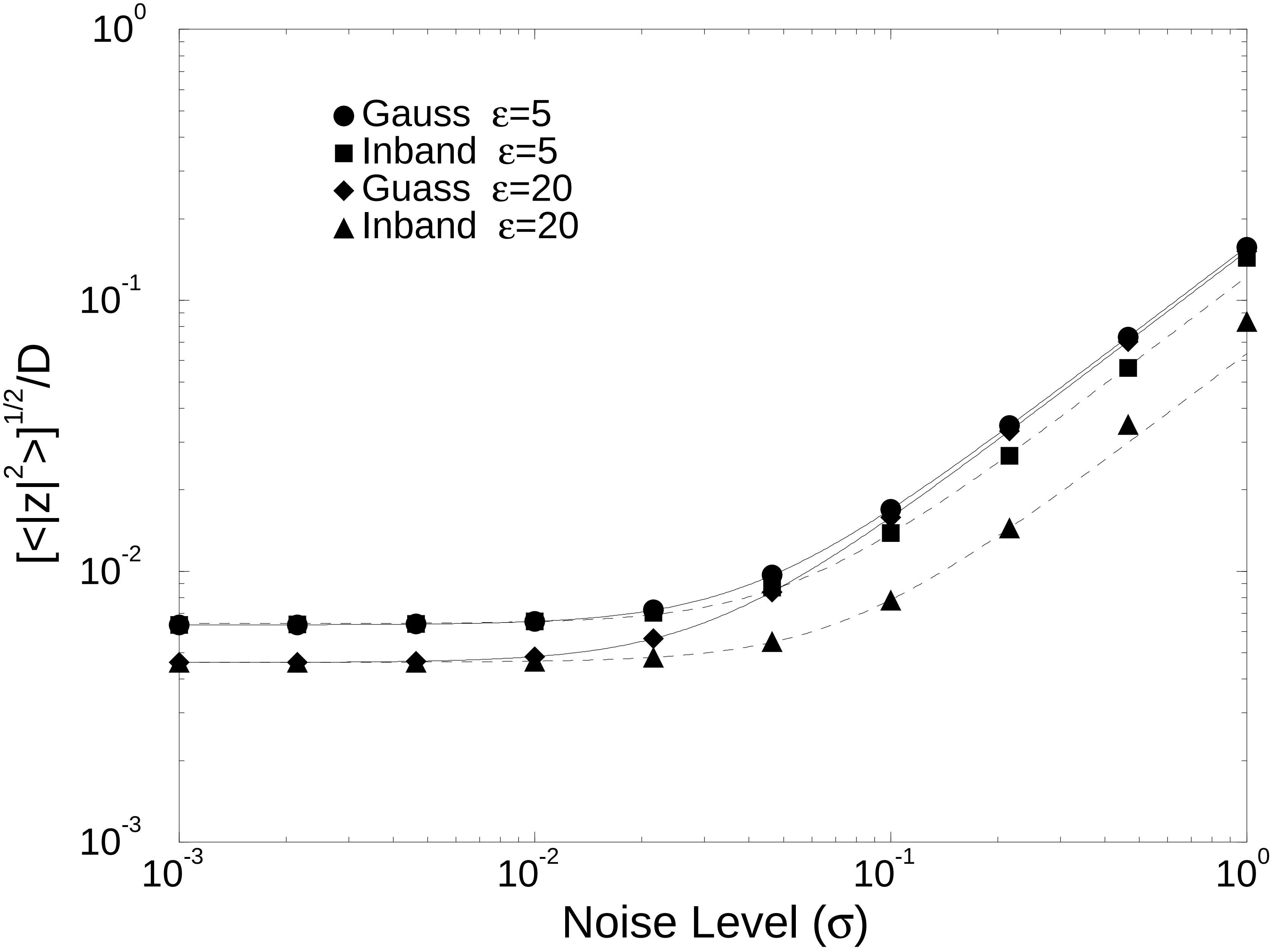

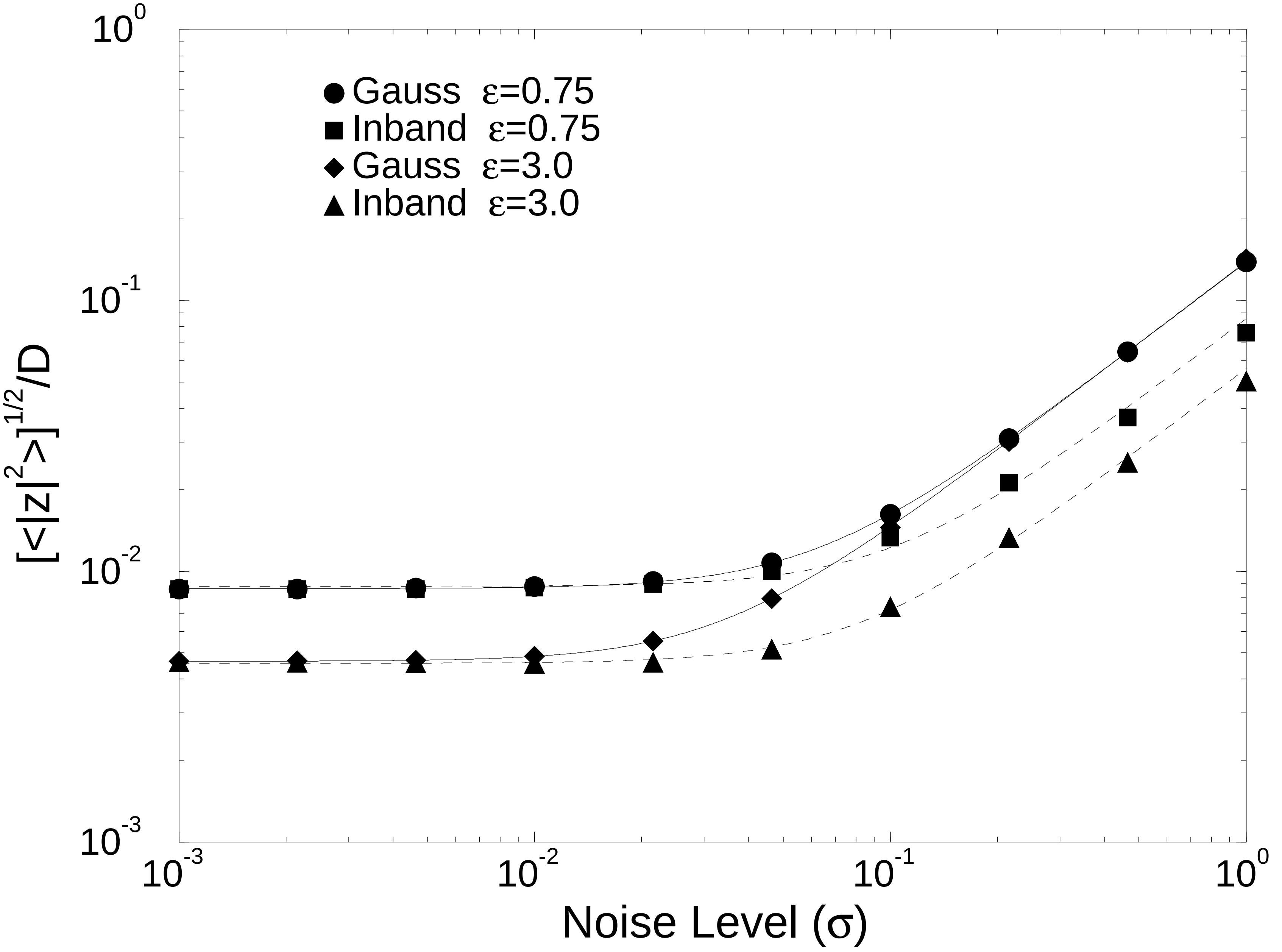

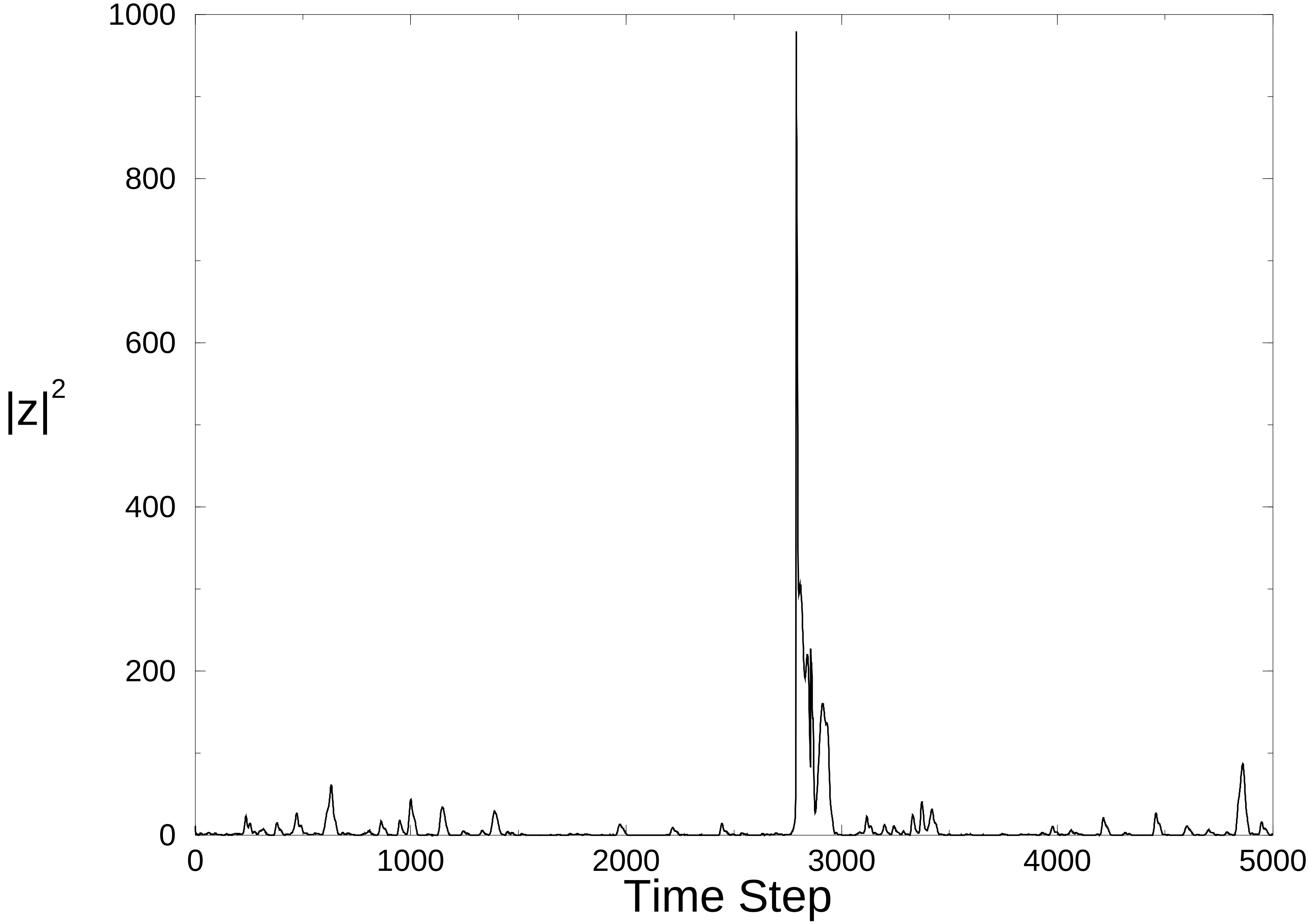

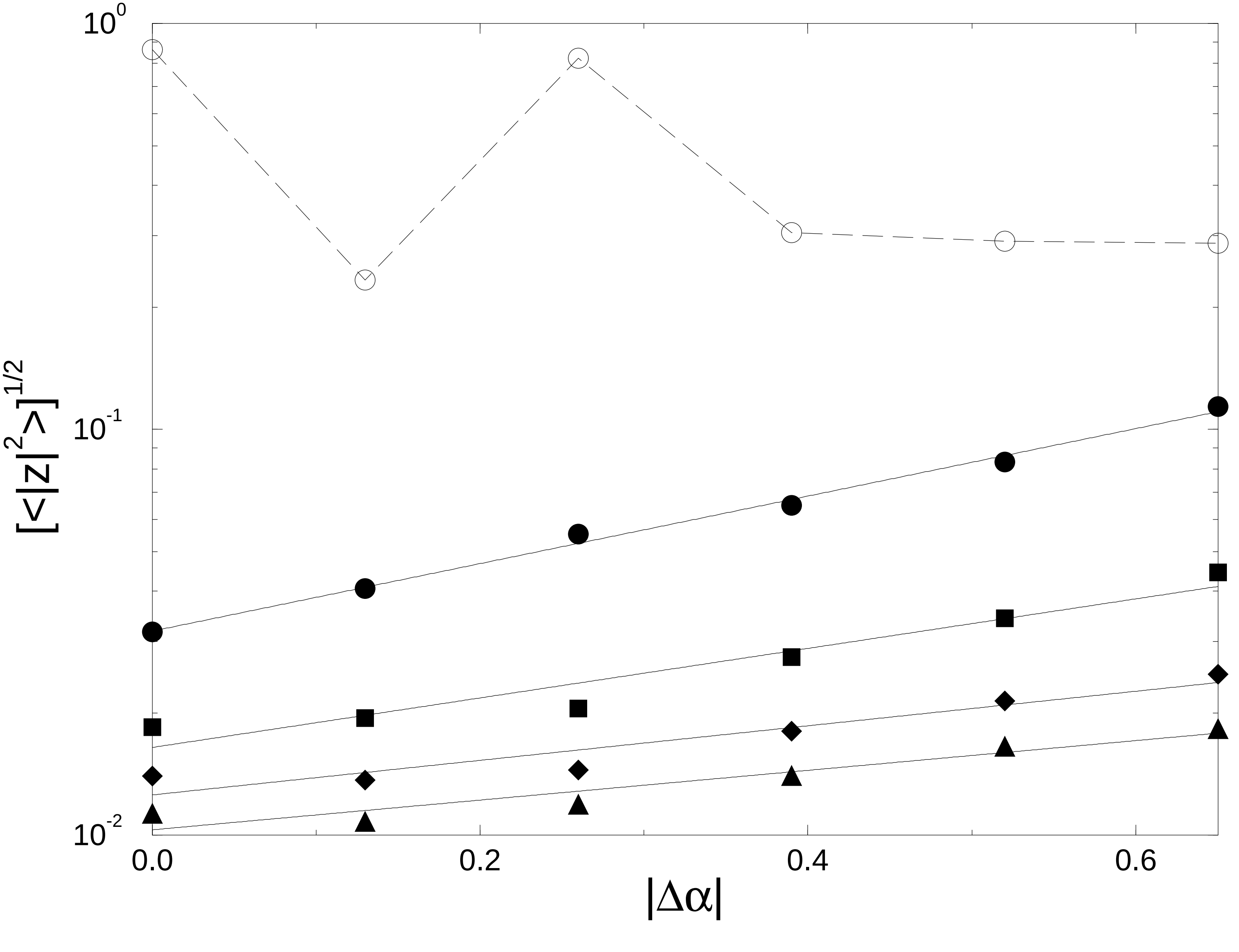

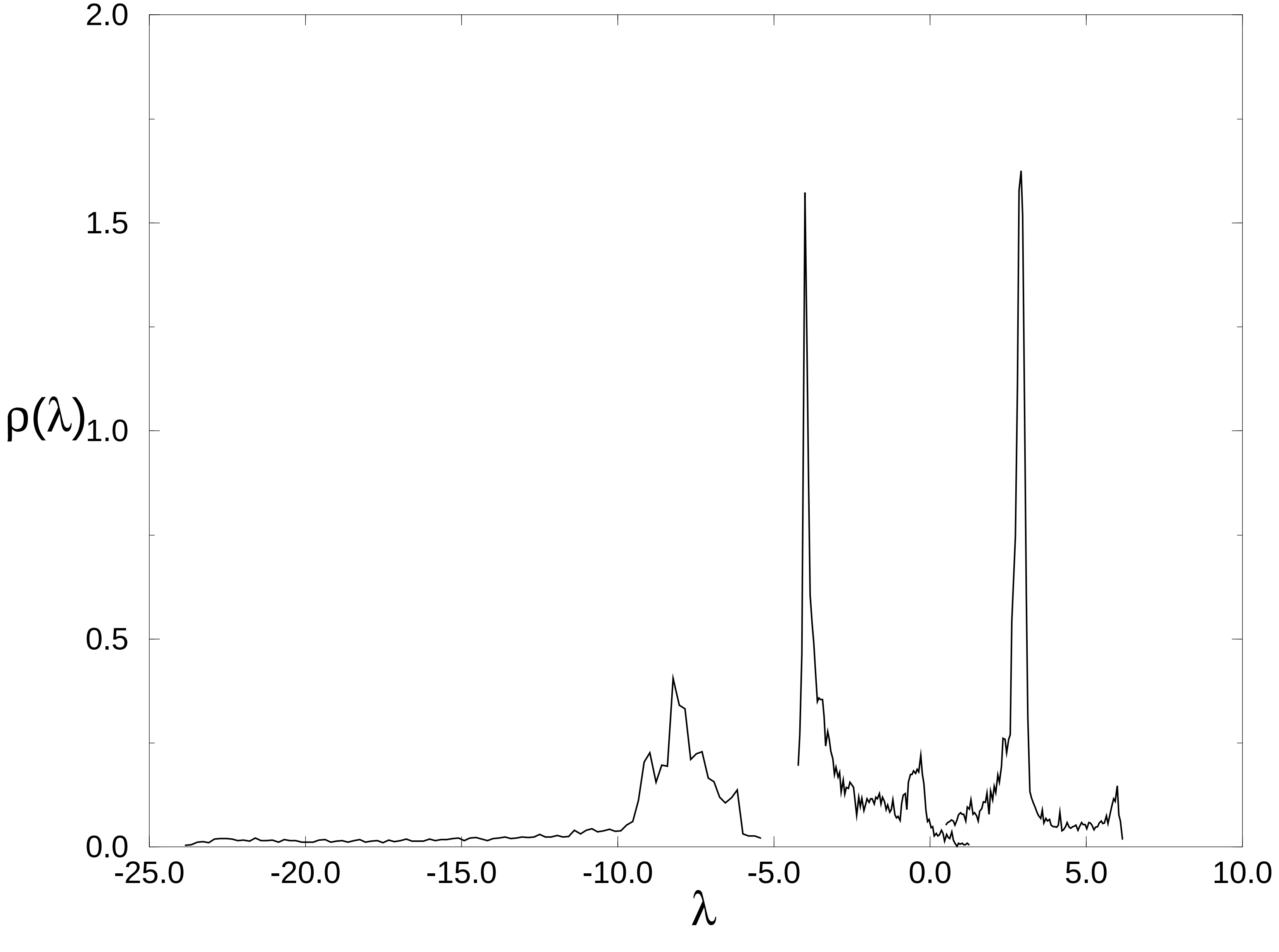

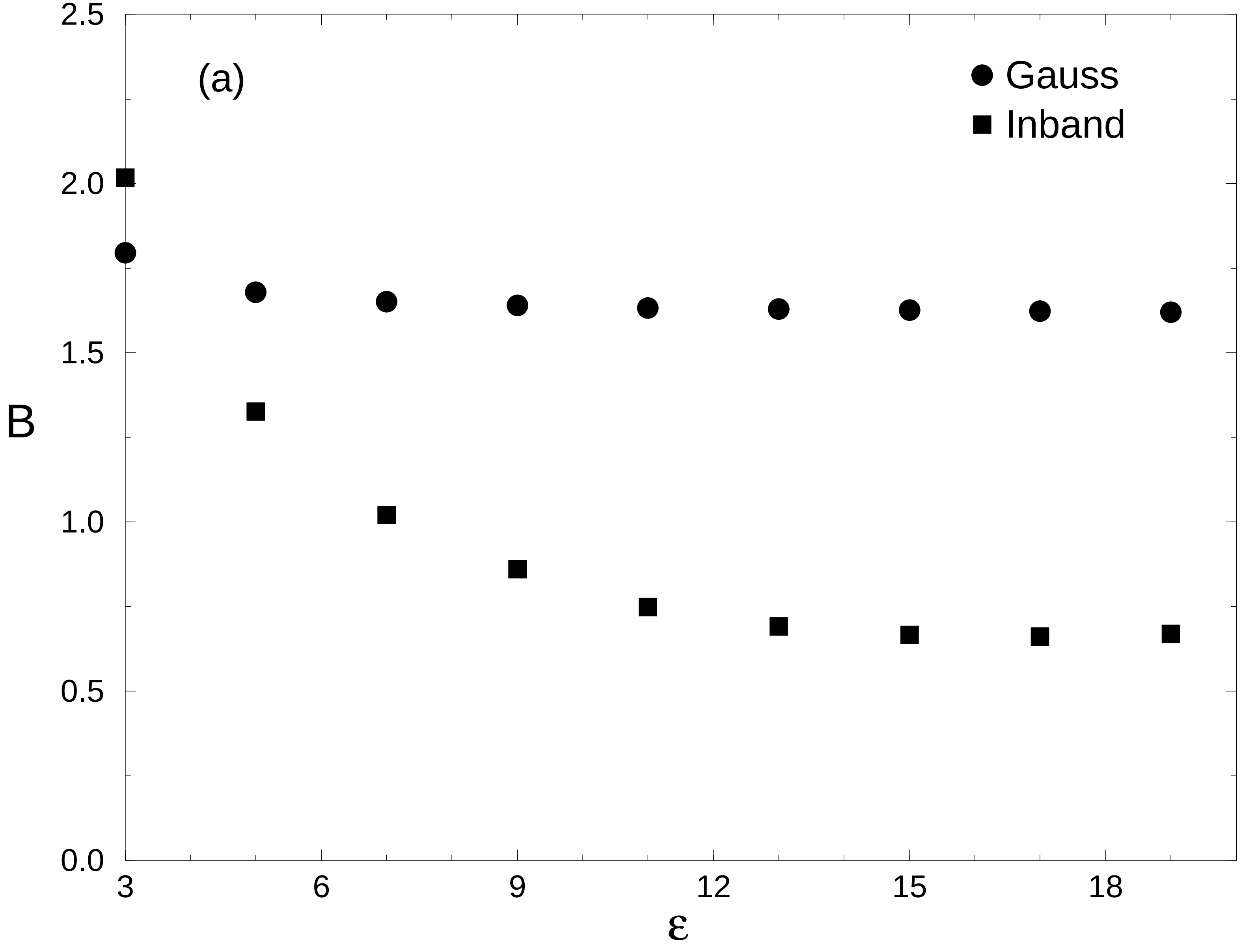

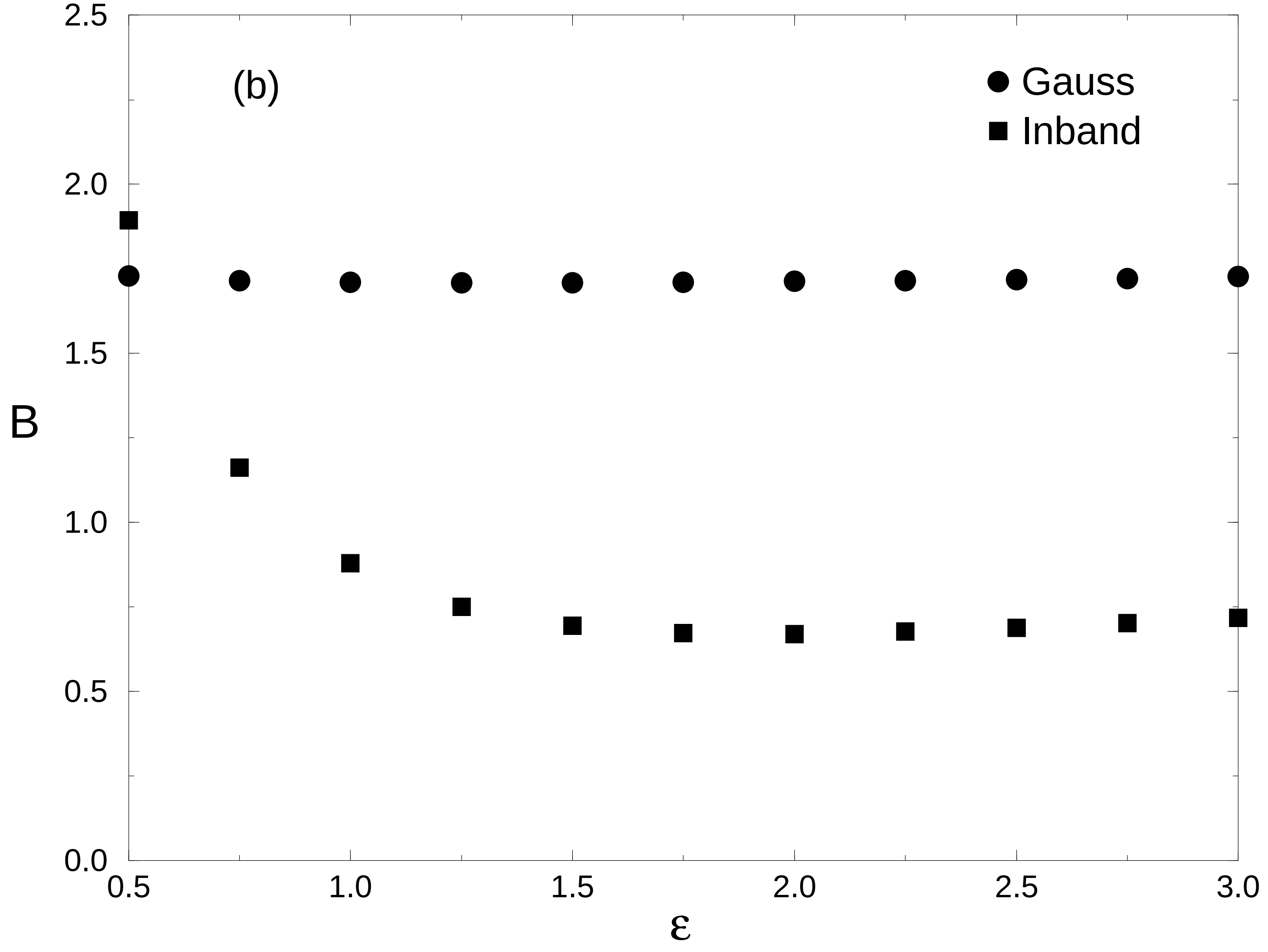

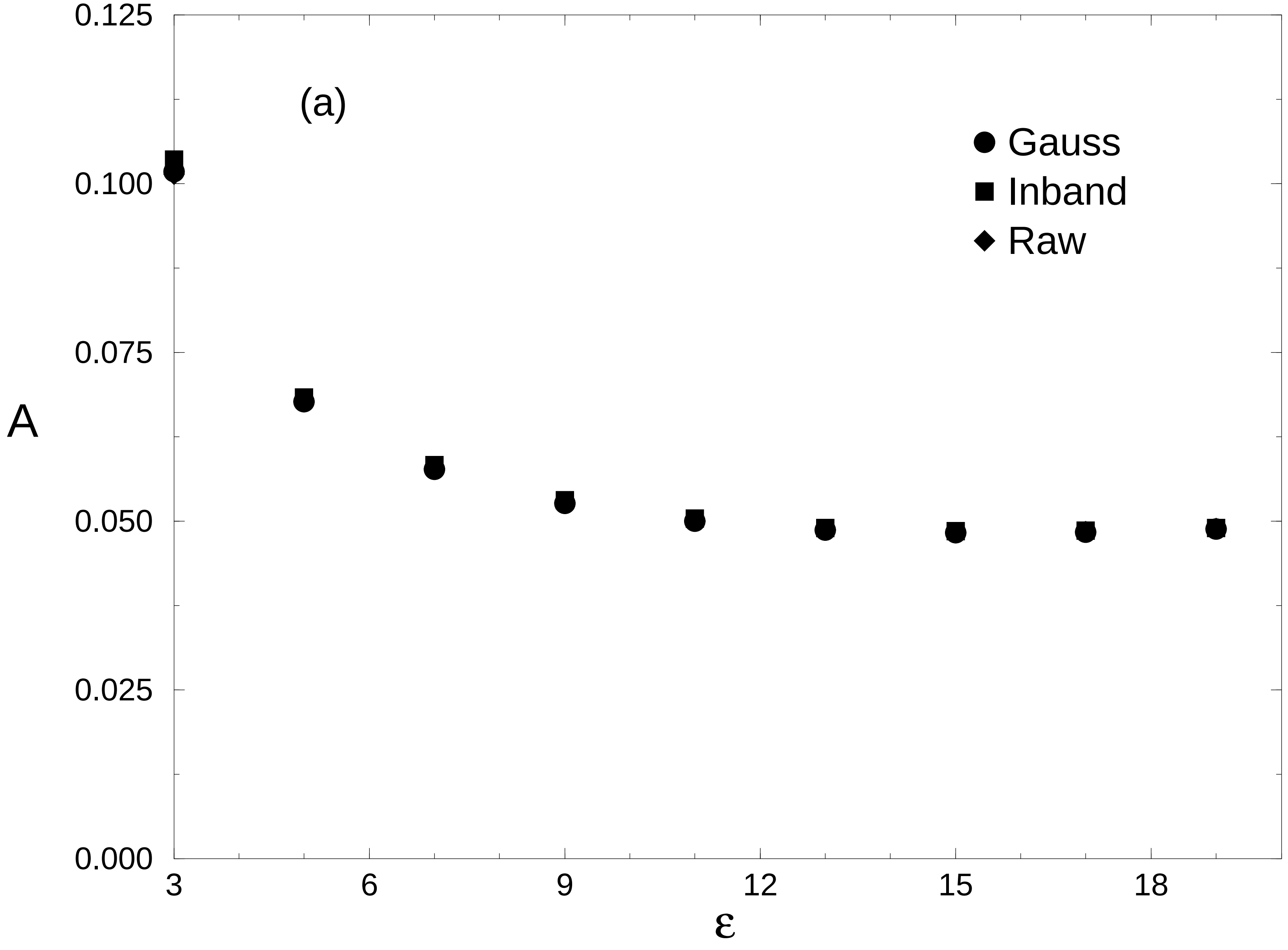

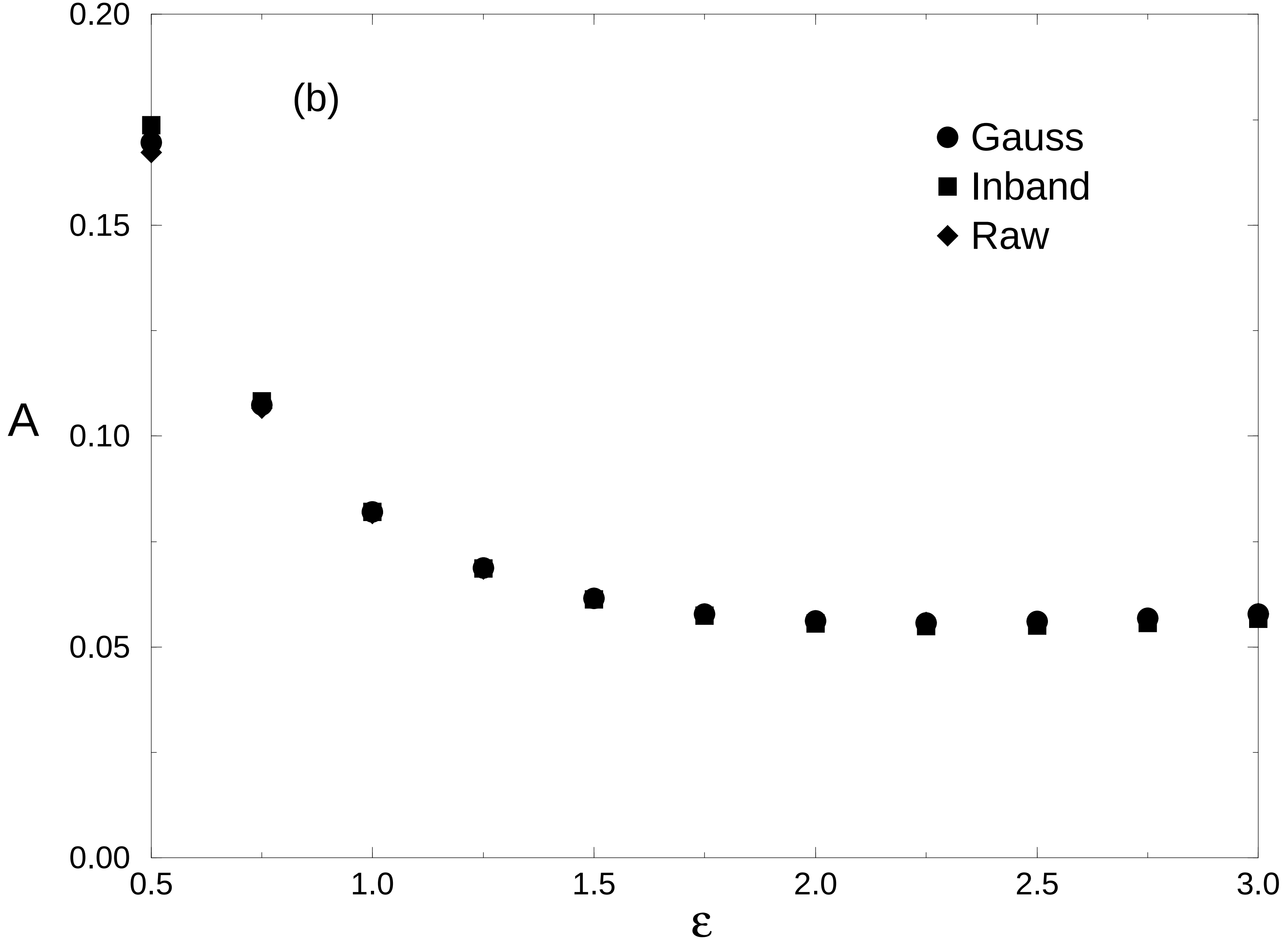

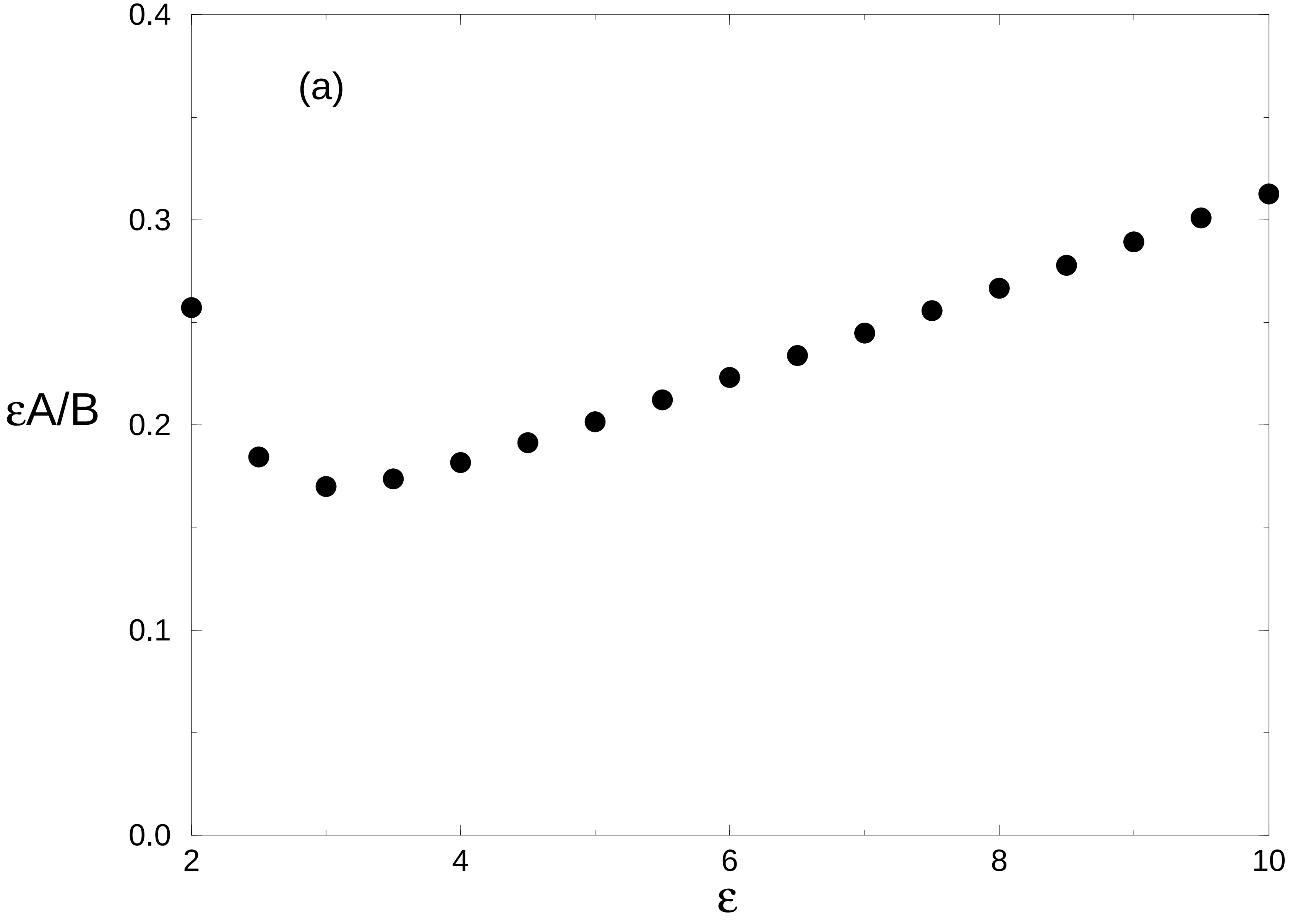

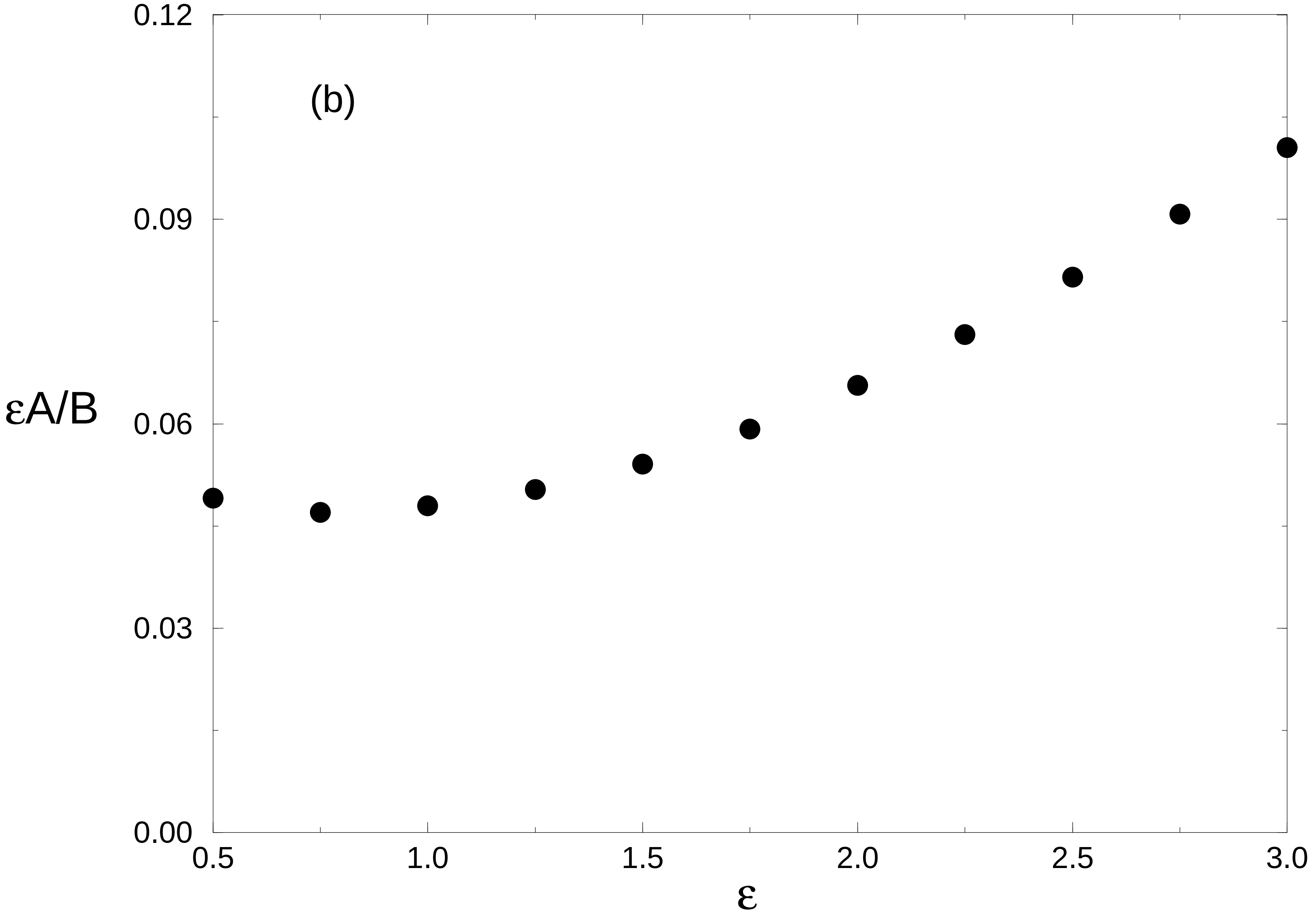

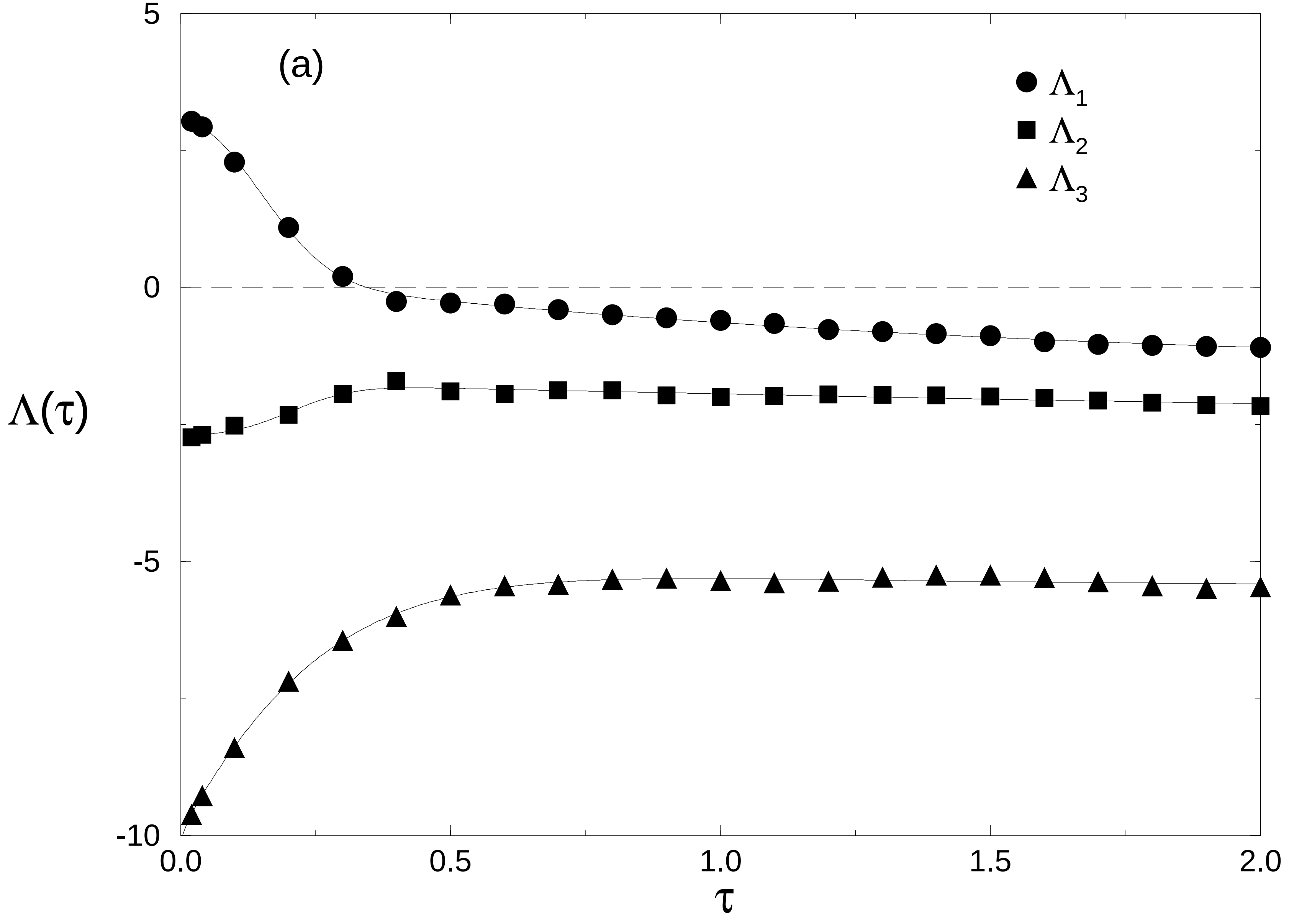

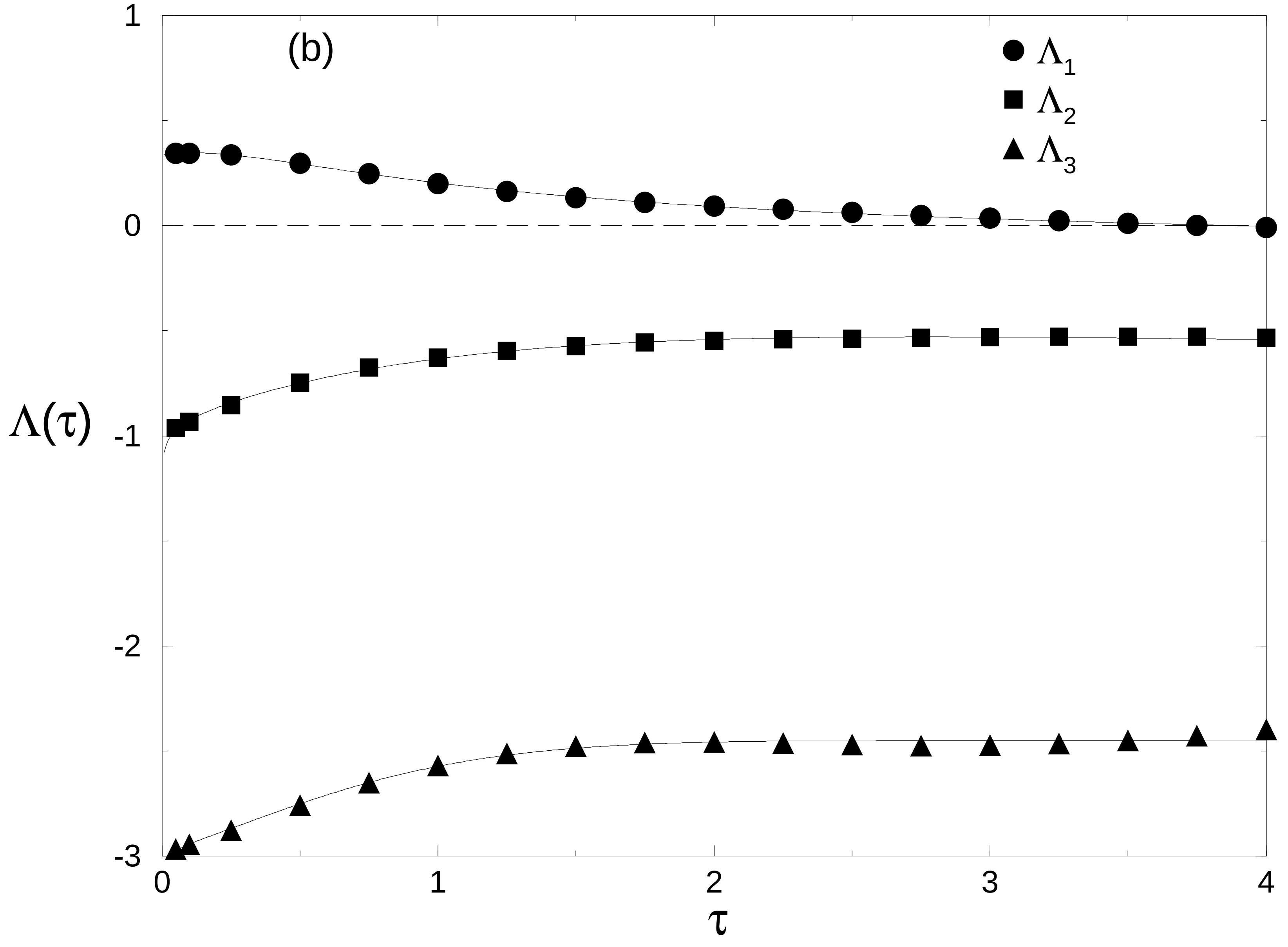

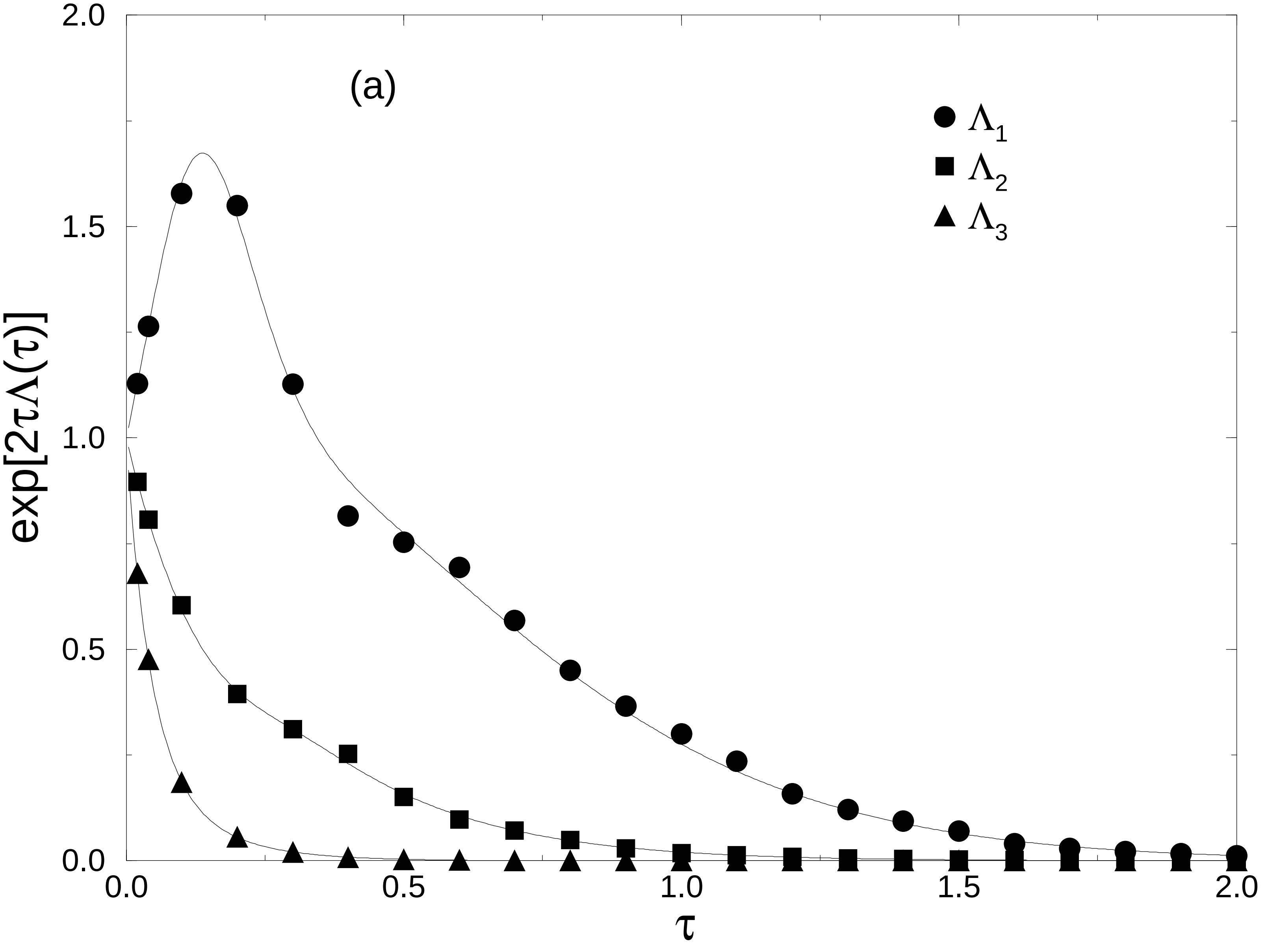

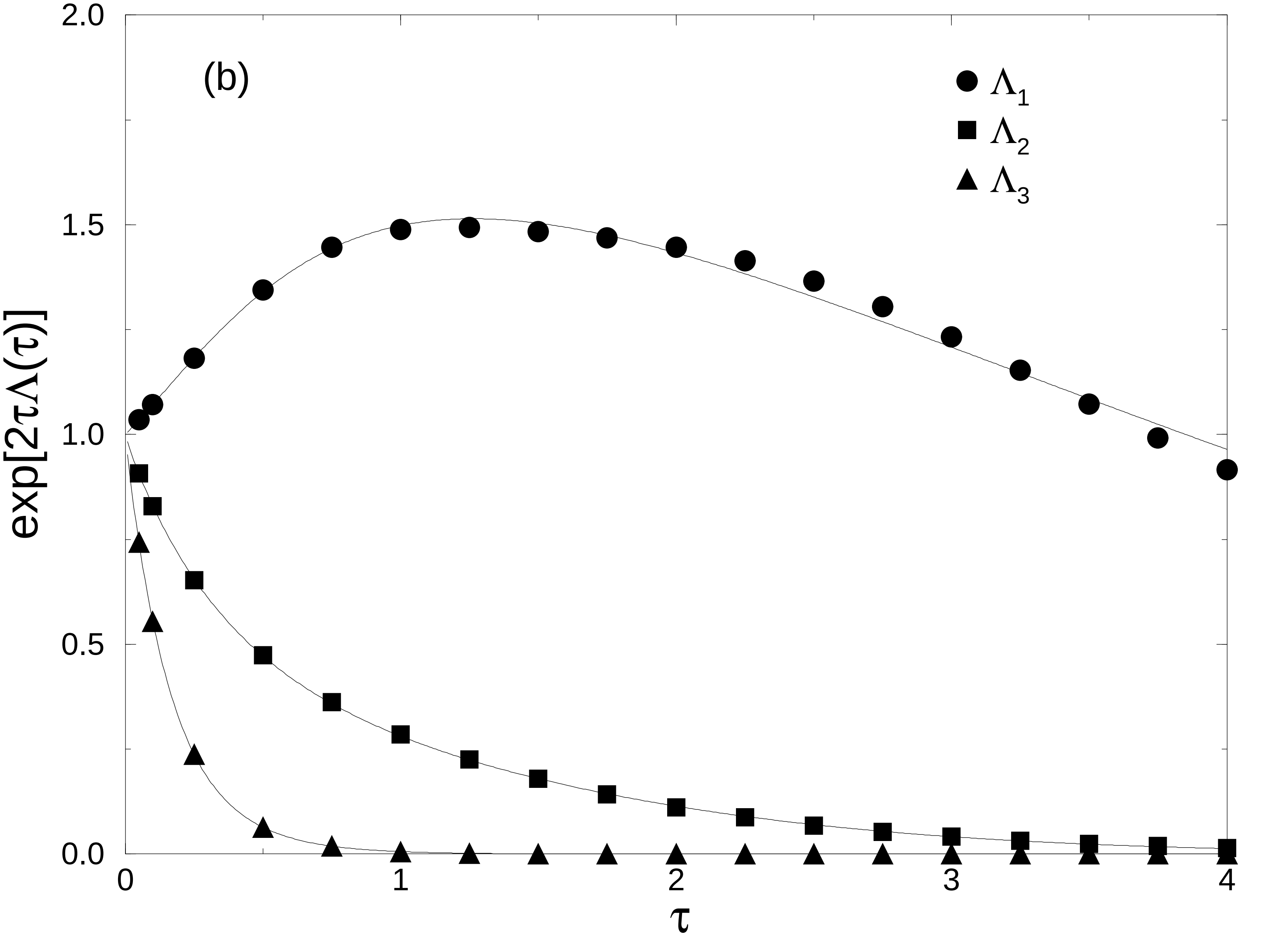

# Synchronization of chaotic systems: The effects of additive noise and drift in the dynamics of the driving

Reggie Brown and Nikolai F. Rulkov
*Institute for Nonlinear Science, University of California, San Diego, La Jolla, CA 92093-0402*

Nicholas B. Tufillaro
*Center for Nonlinear Studies, Los Alamos National Laboratory, Los Alamos, New Mexico 87545*

We examine the effect of additive noise and drift in the dynamics of a chaotic driving signal on the synchronization of a chaotic response systems. Simple scaling laws associated with the synchronization deviation level under these types of contamination are presented. Time series used as the driving signals are experimentally measured from an electronic circuit and a mechanical system (a vibrating wire). The response systems are models which were obtained by fitting an ordinary differential equation to time series data. Possible relevance of this work to non-destructive testing, system identification, and communications is discussed. Finally, we present some results regarding the relationship between the synchronization deviation level and the coupling strength.

## I. INTRODUCTION

Synchronization between two chaotic systems has received considerable attention since it was discussed by Fujisaka and Yamada (FY) and demonstrated by Pecora and Carroll (PC) [1,2]. Many of the papers that have appeared involve theoretical analysis using simple known models [3–6]. In most of these papers synchronization of chaos has been associated with identical, chaotic in time, behavior brought about by the coupling of two or more identical systems. Necessary and sufficient conditions for this sort of synchronization have been discussed and the relationships between various Lyapunov exponents have been investigated [7–10]. Several authors have investigated the relationship between synchronization and control [11,12]. Synchronization has also been demonstrated in a variety of experimental settings [12–15]. Today, it is known that there are many different methods for coupling systems together which can result in synchronous chaotic behavior [16,17]. In an earlier paper we presented results where two different synchronization schemes were employed to check the accuracy of models build from time series data [18].

The discussion in this paper will center around the type of synchronization discussed by most of the authors cited above. Namely, two *identical* chaotic systems that are coupled in a drive/response manner and which exhibit motions that are chaotic and *identical* in time. Unless otherwise specified we will use the term synchronization to mean this form of identical synchronous motion of identical systems. Although identical synchronous motion of identical systems is important it is, none the less, a special case. The general case of synchronization between non-identical systems has been theoretically investigated by Afraimovich, Verichev, and Rabinovich [3] as well as Rulkov et. al. [19]. It has also been experimentally observed by Rulkov and co-workers [13,19]. The general case of synchronization is beyond the scope of this paper and will not be discussed.

For synchronization (i.e. identical synchronous motion of identical systems) any difference between the driving and response systems will break the symmetry between the two systems. Therefore, the behavior of the two systems will no longer be identical. This paper will examine two general questions regarding the appearance of deviations from identical synchronous motion. The first question asks: (1) How will additive noise in the driving signal effect synchronization between the two systems? The second question we examine is more subtle and centers on the fact that no two devices are ever *exactly* the same. This implies that the dynamics of the driving system will differ from the dynamics of the response system (of course, for many situations this difference will be very small). The second question asks: (2) How will differences between the dynamics of the driving and response systems effect synchronization between the two systems? Similar questions have been addressed by Pyragas [12] who stabilized unstable periodic orbits by using a particular form of periodic driving, and by Breeden [20] and Vassiliadis [21] who performed numerical experiments on discrete time systems.

The research we will present attempts to synchronize numerical models to experimentally measured time series data. We will assume that the experimental system is the driving system and that its dynamics are known only through time series measurements. The model is the response system and is assumed to have been constructed from time series measurements taken from the drive system.

The following example serves to illustrate the questions we will address in this paper. Let the unknown dynamics of the driving system in the "working phase space" be represented by

$$\frac{d\mathbf{x}}{dt} = \mathbf{G}(\mathbf{x}), \tag{1}$$

and let the dynamics of the model in the same phase space be represented by

$$\frac{d\mathbf{x}}{dt} = \mathbf{F}(\mathbf{x}). \tag{2}$$

The working phase space is the one where we are modeling the global dynamics of the system that generated the time series. Usually the working phase space is obtained by reconstructing the attractor for the dynamics from the time series. We rely on various forms of the embedding theorem [22,23] to infer the existence of Eq. (1) in the working phase space.

Figure 1 is a pictorial example of an attempt to synchronize a model, Eq. (2), to a time series from Eq. (1). To generate this picture an experimentally measured time series from an electronic circuit (see Section III) was used to construct a model of the dynamics of the circuit. The procedure used to construct the model has been previously discussed [18,24].

Next, we recorded $\mathbf{x}$, a second trajectory from the circuit. Equation (2) and $\mathbf{x}$ were used to generate two new trajectories, $\mathbf{y}$ and $\mathbf{w}$. The trajectory, $\mathbf{w}$, results when the first point in $\mathbf{x}$ is used as an initial condition and Eq. (2) is integrated forward in time without driving. The trajectory, $\mathbf{y}$ results when Eq. (2) is driven by $\mathbf{x}$ (using the driving method discussed in Section II).

The lower curve in Fig. 1 is the square of the Euclidean distance between the trajectory of the driving time series and the trajectory of the driven model, $|\mathbf{z}|^2 = |\mathbf{x}-\mathbf{y}|^2$. The top curve is the square of the Euclidean distance between the trajectory of the driving time series and the trajectory of the non-driven model, $|\mathbf{z}|^2 = |\mathbf{x}-\mathbf{w}|^2$. The fact that the distances shown in the lower curve are small when compared to those in the upper curve indicates that the deviations from identical synchronous motion are small and the trajectory of the driven model does not deviate much from the driving trajectory. If $0 < |\mathbf{z}|^2 \ll 1$ then we will say that the drive and response systems are almost synchronized. Heuristically speaking, the level of the lower curve indicates the quality of synchronization between the drive and response systems.

In the complete absence of noise and modeling errors (i.e., $\mathbf{F} = \mathbf{G}$) we expect $|\mathbf{z}|^2 = 0$, (i.e., the drive and response systems are synchronized). As the systems approach synchronization the level of the lower curve in Fig. 1 approaches $-\infty$. For physical devices and model equations this will never happen since noise is always present in the driving signal and there are always modeling errors. Hence, a physical devices and a model can only be almost synchronized. Similarly, two physical devices can only be almost synchronized since no two devices are ever *identical.* As the size of the noise in the driving signal and/or the size of the modeling errors increase we expect the level of the lower curve in Fig. 1 to rise. Understanding the behavior of this rise as a function of noise and modeling errors will answer questions (1) and (2).

In this paper we have examined two different types of additive noise in the driving signal. The first type of noise is white Gaussian noise. The second type of noise (called inband noise) is constructed to have the same power spectrum as the clean driving signal. Inband noise is used because it represents a type of noise that is very hard to remove using standard filters based on Fourier transforms.

We close this section with an outline of the remainder of this paper. In Section II we derive theoretical predictions for the effects of noise and modeling errors on synchronization. In addition, we discuss how, in the no noise limit, the level of the lower curve in Fig. 1 can be used as a measure of modeling errors. Section III presents the results of numerical experiments we have conducted using models obtained from the analysis of experimentally measured data. The data comes from an electronic circuit and a mechanical system (a vibrating wire). Finally, in Section IV we summarize our results and present some conclusions. We close Section IV, and motivate our research, by presenting various applications where our study of synchronization may be useful. These applications involve non-destructive testing, communications, and system identification.

This paper also contains three appendices. In the appendices we present a partial analysis of the effect of the coupling strength on synchronization. We believe that this discussion is the first analytical discussion of this effect for the type of coupling we have investigated.

## II. THEORETICAL CONSIDERATIONS

In this section we present part of the theoretical analysis behind our study of the deviations from synchronization (which we call the synchronization deviation level, or simply the deviation level) in the presence of noise in the driving signal and/or drift in the dynamics of the driving signal. After introducing notation the basic equations needed to examine the deviations are derived. This Section closes with a definition of the synchronization deviation level and expressions for the constant terms that appear in this definition.

The method we will use to synchronize our models to the driving system is a modification of the one proposed by FY [1] and has been studied in a previous paper [18]. A synchronization scheme that is similar to ours has been implemented by Newell et. al. [14] on pairs of diode resonator circuits, by Rulkov and Volkovskii [25] on electronic circuits with one way drive/response style coupling, as well as many of the other papers cited in the introduction.

We are assuming that the time series used for the driving system consists of $d$ dimensional vectors that live in the working phase space. How one obtains these vectors, either by direct measurement or some form of phase space reconstruction, has been discussed elsewhere [23]. The time evolution of these vectors traces out an attractor (we consider chaotic attractors) which represents the temporal dynamics of the driving system.

In order to establish notation assume that, in the working phase space, Eq. (1) represents the true dynamics of the device that is producing the driving signal. In practice, the exact from of the vector field, $\mathbf{G}$, is usually unknown. In what follows $(\mathbf{x}+\sigma\hat{\mathbf{u}})$ is used to indicate the driving time series. The vector $\mathbf{x}$ represents the clean time series from $\mathbf{G}$ while $\sigma\hat{\mathbf{u}}$ represents additive noise in the experimentally recorded time series. $\sigma$ indicates the size of the noise while the time series $\hat{\mathbf{u}}$ is of unit size. The noise term is often associated with errors in measurements of the driving signal. For example, these errors could be the result of a faulty measuring device, or background noise that is being measured along with the signal by a good measuring device.

The model, denoted by Eq. (2), is constructed from a time series which is not the same as the driving time series (although they may have the same source). The details of how we construct ODE models from a time series have been presented in Refs. [18,24]. Reference [18] also demonstrates that by coupling $\mathbf{F}$ to a noisy time series from $\mathbf{G}$ via

$$\frac{d\mathbf{y}}{dt} = \mathbf{F}(\mathbf{y}) - \mathbf{E}\cdot[\mathbf{y}-(\mathbf{x}+\sigma\hat{\mathbf{u}})], \tag{3}$$

it is possible to almost synchronize the dynamics of $\mathbf{F}$ to $\mathbf{x}$ (i.e., the distance between $\mathbf{y}$ and $\mathbf{x}$ is small).

The matrix $\mathbf{E}$ represents the coupling between $\mathbf{y}$ and the experimentally recorded time series. For the cases we will study $\mathbf{E}$ has only one nonzero element. That element lies on the diagonal and is $E_{\beta\beta} = \epsilon$ when the $\beta$ component of $(\mathbf{x}+\sigma\hat{\mathbf{u}})$ is used as the driving signal. Within some range of values for $\epsilon$ the kind of dissipative coupling shown in Eq. (3) can result in almost identical chaotic oscillations. As we will show, any lack of complete synchronization between $\mathbf{x}$ and $\mathbf{y}$ in Eq. (3) is caused by modeling errors and noise.

Obviously, if $\epsilon$ is below some critical threshold, $\epsilon_c$, then synchronization will not occur. The $\epsilon \to \infty$ limit is considered in detail in Appendix A. Note that synchronized motion, or almost synchronized motion, may only be possible within some finite range of values for $\epsilon$. For these situations using a value of $\epsilon$ that is too small *or* too large will not produce synchronous, or almost synchronous motion. An example of this for a case using periodic driving is discussed in the paper by Pyragas [12]. An example of this behavior for chaotic driving can be found in Ref. [18].

In our numerical experiments the coupling occurs only in the last component of $\mathbf{F}$, although we could have used other components or combinations of other components [16,18].

As a definition we say that the model of the device, $\mathbf{F}$, is synchronized to the time series, $\mathbf{x}$, if $\mathbf{x}=\mathbf{y}$ *and* $\mathbf{F}=\mathbf{G}$ for all time greater than some transient, $t_0$. Clearly, when $\mathbf{F}\neq\mathbf{G}$ and/or in the presence of noise synchronization is not possible. Let $\mathbf{z}=\mathbf{y}-\mathbf{x}$ denote the deviations between $\mathbf{y}$ and the clean driving signal, $\mathbf{x}$. If the deviations are small, $0<|\mathbf{z}|^2\ll 1$, then the linearized time evolution of $\mathbf{z}$ is given by

$$\frac{d\mathbf{z}}{dt} = [\mathbf{DF}(\mathbf{x})-\mathbf{E}]\cdot\mathbf{z} + \sigma\mathbf{E}\cdot\hat{\mathbf{u}} + \Delta\mathbf{G}(\mathbf{x}), \tag{4}$$

where $\Delta\mathbf{G}=\mathbf{F}-\mathbf{G}$ denotes the difference between the model and the true dynamics of the driving system.

It is important to remember that $\Delta\mathbf{G}$ has two potential sources, which are assumed to be unrelated to measurement errors in the time series. The first source of $\Delta\mathbf{G}$ is associated with errors in our attempt to model the unknown vector field, $\mathbf{G}$. For any real situation $\mathbf{F}$ is never exactly equal to $\mathbf{G}$ so it is reasonable to expect that this source of error will always be present. The second source comes about if the dynamics of the driving signal is different from the dynamics that produced the time series used to make the model.

To analytically isolate these two sources assume that the time series used to generate the model comes from the vector field $\mathbf{G}'$, while the time series used to drive the model comes from the vector field $\mathbf{G}$. We also assume that $\mathbf{G}$ and $\mathbf{G}'$ are related by some small change in $\mathbf{p}$, the parameters of the driving system (i.e., $\mathbf{G}\simeq\mathbf{G}'+(\partial\mathbf{G}'/\partial\mathbf{p})\cdot\delta\mathbf{p}$). Under these conditions the analysis that leads to Eq. (4) results in

$$\Delta\mathbf{G}(\mathbf{x}) \simeq \Delta\mathbf{G}'(\mathbf{x}) + \left(\frac{\partial}{\partial\mathbf{p}}\Delta\mathbf{G}'(\mathbf{x})\right)\cdot\delta\mathbf{p}, \tag{5}$$

where $\Delta\mathbf{G}' = \mathbf{F} - \mathbf{G}'$. The first term is associated with modeling errors while the second term is associated with drift in the dynamics of the driving system.

Equations (4) and (5) are the principle evolution equations for coupled systems in the vicinity of synchronized chaotic motion. They describe the evolution of $\mathbf{z}$, the vector associated with the deviations between $\mathbf{F}$, our model of the dynamics and $\mathbf{x}$, the clean time series from the driving device. By knowing the behavior of $\mathbf{z}$ we will be able to answer the two questions asked in the introduction.

In order to formally integrate Eq. (4) consider the case of no noise in the driving time series, $\sigma = 0$, and perfect modeling of the dynamics of the driving signal, $\Delta\mathbf{G}(\mathbf{x}) = \mathbf{0}$. Under these conditions Eq. (4) becomes

$$\frac{d\mathbf{z}}{dt} = [\mathbf{DF}(\mathbf{x}) - \mathbf{E}] \cdot \mathbf{z},$$

which represents the homogeneous portion of Eq. (4). This equation is a linear ODE with time dependent coefficients whose formal solution is

$$\begin{aligned} \mathbf{z}(t) &= \exp\left[\int_{t_0}^{t} [\mathbf{DF}(r) - \mathbf{E}]\, dr\right] \cdot \mathbf{z}(t_0) \\ &= \mathbf{U}(t, t_0) \cdot \mathbf{z}(t_0), \end{aligned} \tag{6}$$

where $\mathbf{DF}(r) = \mathbf{DF}[\mathbf{x}(r)]$, and the second equation defines $\mathbf{U}(t, t_0)$. The matrix $\mathbf{U}(t, t_0)$ is the evolution operator that evolves the initial condition, $\mathbf{z}(t_0)$, forward in time from $t_0$ to $t$ in the presence of coupling and the absence of noise and modeling errors. This operator satisfies the initial condition $\mathbf{U}(t_0, t_0) = \mathbf{1}$ where $\mathbf{1}$ is the identity. Since we have assumed that the time evolution is stable (the system synchronizes) we know that $\mathbf{U}(t, t_0)$ shrinks $|\mathbf{z}(t_0)|$ to zero exponentially fast as $t \to \infty$. The rate of decrease is controlled in a nontrivial fashion by the coupling $\epsilon$ [1,17].

In order to obtain the general solution to Eq. (4) we must add a specific solution to the homogeneous solution given by Eq. (6). To obtain a specific solution to Eq. (4) in the presence of noise and modeling errors change variables from $\mathbf{z}$ to $\mathbf{w}$ via $\mathbf{z}(t) = \mathbf{U}(t, t_0) \cdot \mathbf{w}(t)$. Inserting this into Eq. (4) gives

$$\frac{d\mathbf{w}}{dt} = \mathbf{U}^{-1}(t, t_0) \cdot [\Delta\mathbf{G}(t) + \sigma\mathbf{E} \cdot \hat{\mathbf{u}}(t)].$$

By formally integrating this equation and making use of $\mathbf{z}(t) = \mathbf{U}(t, t_0) \cdot \mathbf{w}(t)$ we finally arrive at the desired expression

$$\mathbf{z}(t) = \mathbf{U}(t, t_0) \cdot \mathbf{z}(t_0) + \int_{t_0}^{t} \mathbf{U}(t, r) \cdot [\Delta\mathbf{G}(r) + \sigma\mathbf{E} \cdot \hat{\mathbf{u}}(r)]\, dr. \tag{7}$$

Equation (7) is a formal solution to Eq. (4) and is one of the principle results of this section. The equation represents the time evolution of the difference between the trajectory given by Eq. (3) and the trajectory of the true system, Eq. (1), when the systems are almost synchronized. Because of the stability of synchronized motion the first term in Eq. (7) can be ignored since it will vanish exponentially fast as $t$ increases. The operator $\mathbf{U}(t, r)$ in the second term in Eq. (7) evolves fluctuations in $\Delta\mathbf{G}$ and $\sigma\hat{\mathbf{u}}$ forward in time. As the fluctuations evolve their magnitude decays due to the properties of the evolution operator, $\mathbf{U}(t, r)$. The integral over $r$ indicates that $\mathbf{z}(t)$ is equal to the fluctuations occurring at time $t$ as well as the sum of the decayed fluctuation that have occurred at previous times.

The bottom curve in Fig. 1 indicates that, while the detailed time behavior of $\log_{10}(|\mathbf{z}|^2)$ is quite complicated, the average behavior is essentially constant. This average can be used to indicate the synchronization deviation level between the clean driving signal and the model. To be specific, we define the synchronization deviation level (the deviation level) by the following time average

$$\left[\langle|\mathbf{z}|^2\rangle_T\right]^{1/2} = \left[\lim_{t\to\infty} \frac{1}{t - t_0} \int_{t_0}^{t} |\mathbf{z}(r)|^2 dr\right]^{1/2}. \tag{8}$$

Equation (8) is one of the principle definitions of this section. The synchronization deviation level indicates the quality of synchronization between the driving and the response systems. The smaller the deviation level the closer we are to complete synchronization.

To formally calculate the time average in Eq. (8) we use Eq. (7) to write $|\mathbf{z}|^2$ as

$$\begin{aligned} |\mathbf{z}(t)|^2 = &\left[\int_{t_0}^{t} \mathbf{U}(t, r) \cdot \Delta\mathbf{G}(r) dr\right] \cdot \left[\int_{t_0}^{t} \mathbf{U}(t, r) \cdot \Delta\mathbf{G}(r) dr\right] \\ &+ 2\sigma\epsilon \left[\int_{t_0}^{t} \eta(r)\, \mathbf{U}(t, r) \cdot \mathbf{V} dr\right] \cdot \left[\int_{t_0}^{t} \mathbf{U}(t, r) \cdot \Delta\mathbf{G}(r) dr\right] \\ &+ (\sigma\epsilon)^2 \left[\int_{t_0}^{t} \eta(r)\, \mathbf{U}(t, r) \cdot \mathbf{V} dr\right] \cdot \left[\int_{t_0}^{t} \eta(r)\, \mathbf{U}(t, r) \cdot \mathbf{V} dr\right], \end{aligned} \tag{9}$$

where $\mathbf{V}$ is the $d$ dimensional vector $\mathbf{V} = [0, \ldots, 0, 1, 0, \ldots, 0]$. The 1 appears as the $\beta$ component of $\mathbf{V}$ if we are using the $\beta$ component of $(\mathbf{x} + \sigma\hat{\mathbf{u}})$ as the driving variable. We have also written the $\beta$ component of $\hat{\mathbf{u}}$ as $\eta$.

The remainder of this section is devoted to an analysis of the three dot products in Eq. (9). Assuming the noise is ergodic allows us to replace time averages by phase space averages, which we denote by brackets, $\langle\rangle$. We also argue that the noise is completely independent of the $\Delta\mathbf{G}$ vector and the evolution operator, $\mathbf{U}(t, t_0)$.

The analysis begins with an examination of the second term in Eq. (9), which is a dot product between two different vectors. Given the assumptions just mentioned the time average of the second term in Eq. (9) becomes

$$\begin{aligned}\langle I_2\rangle_T &= 2\sigma\epsilon \left\langle \left[\int_{t_0}^{t} \eta(r)\mathbf{U}(t, r) \cdot \mathbf{V} dr\right] \cdot \left[\int_{t_0}^{t} \mathbf{U}(t, r) \cdot \Delta\mathbf{G}(r) dr\right]\right\rangle \\ &= 2\sigma\epsilon \left[\int_{t_0}^{t} \langle\eta(r)\rangle \langle\mathbf{U}(t, r) \cdot \mathbf{V}\rangle\, dr\right] \cdot \left\langle \left[\int_{t_0}^{t} \mathbf{U}(t, r) \cdot \Delta\mathbf{G}(r) dr\right]\right\rangle .\end{aligned}$$

The noise we have utilized in Section III has zero mean. For zero mean noise $\langle\eta(r)\rangle = 0$, and the time average of the second term in Eq. (9) vanishes.

We now turn our attention to the last term in Eq. (9). Because this term is the dot product of a vector with itself it is positive semi-definite. It will be useful to begin our analysis by rewriting this term as

$$I_3(t, t_0) = (\sigma\epsilon)^2 \int_{t_0}^{t}\int_{t_0}^{t} \eta(r)\,\eta(r')\,\left[\mathbf{U}(t, r) \cdot \mathbf{V}\right] \cdot \left[\mathbf{U}(t, r') \cdot \mathbf{V}\right]\, dr'\, dr.$$

The domain of integration for the double integral is a square. The fact that the integrand is symmetric with respect to $r$ and $r'$ can be used to rewrite $I_3$ as the sum of two integrals, one along the diagonal of the square, $r = r'$, and one over the triangle defined by $r > r'$. Furthermore, if we change variables from $r'$ to $s = r - r'$ and equate time averages with ensemble averages then the time average of $I_3(t, t_0)$ becomes

$$\begin{aligned}\langle I_3\rangle_T &= (\sigma\epsilon)^2 \int_{t_0}^{t} \langle\eta^2(r)\rangle \langle[\mathbf{U}(t, r) \cdot \mathbf{V}] \cdot [\mathbf{U}(t, r) \cdot \mathbf{V}]\rangle\, dr \\ &+ 2(\sigma\epsilon)^2 \int_{t_0}^{t}\int_{0^+}^{r} \langle\eta(r)\eta(r - s)\rangle \langle[\mathbf{U}(t, r) \cdot \mathbf{V}] \cdot [\mathbf{U}(t, r - s) \cdot \mathbf{V}]\rangle\, ds\, dr,\end{aligned} \tag{10}$$

where the assumed independence of the noise and the dynamics has been used. The plus sign superscript in the last integral indicates that the lower limit of this integral is understood to be arbitrarily close to zero but not equal to zero. This implied limit present no difficulty as long as the integrand does not diverge along the diagonal.

The ensemble averages are autocorrelation functions. The one associated with the noise, $\langle\eta(r)\eta(r - s)\rangle$, often appears in the analysis of Brownian motion [26]. We will assume that the noise is stationary (an assumption that is valid for the type of noise we have investigated). This implies that the ensemble average removes the explicit $r$ dependence from the noise autocorrelation, thereby reducing it to only a function of $s$, $\langle\eta(r)\eta(r - s)\rangle = k(s)$.

Notice that the integral over $s$ in Eq. (10) only acts on $\mathbf{U}(t, r - s)$ and $k(s)$. This allows us to simplify the notation by rewriting this integral as

$$\begin{aligned}\int_{0^+}^{r} k(s)\mathbf{U}(t, r - s) \cdot \mathbf{V} ds &= k(0)\mathbf{U}(t, r) \cdot \int_{0^+}^{r} \frac{k(s)}{k(0)}\mathbf{U}(r, r - s) \cdot \mathbf{V} ds \\ &= k(0)\mathbf{U}(t, r) \cdot \mathbf{B}[\mathbf{x}(r)],\end{aligned} \tag{11}$$

where the second equality serves to define the vector $\mathbf{B}[\mathbf{x}(r)]$. Using Eq. (11) we can rewrite Eq. (10) as

$$\begin{aligned}\langle I_3\rangle_T &= (\sigma\epsilon)^2 k(0) \left[\int_{t_0}^{t} \left\langle |\mathbf{U}(t, r) \cdot \mathbf{V}|^2\right\rangle dr + 2\int_{t_0}^{t} \langle[\mathbf{U}(t, r) \cdot \mathbf{V}] \cdot [\mathbf{U}(t, r) \cdot \mathbf{B}(r)]\rangle\, dr\right] \\ &= \sigma^2 B^2,\end{aligned} \tag{12}$$

where the last equation serves to define $B$. The form of the integral, $\langle I_3\rangle_T$, shown in Eq. (12) is exact. $B$ is a nontrivial function of $\epsilon$ and the type of noise, but is *not* a function of $\sigma$ or the modeling errors, $\Delta\mathbf{G}$. Therefore, $\langle I_3\rangle_T$ is a quadratic function of the size of the noise.

We now turn our attention to the first term in Eq. (9). The analysis of this integral will parallel the one used for $I_3(t, t_0)$, hence, we will only briefly discuss the various mathematical manipulations. Begin by rewriting the first integral in Eq. (9) as

$$I_1(t,t_0) = \int_{t_0}^{t}\int_{t_0}^{t} [\mathbf{U}(t,r)\cdot\Delta\mathbf{G}(r)]\cdot[\mathbf{U}(t,r')\cdot\Delta\mathbf{G}(r')]\, dr'\, dr. \tag{13}$$

Taking advantage of symmetry and changing variables from $r'$ to $s = r - r'$ produces

$$\begin{aligned} I_1(t,t_0) = &\int_{t_0}^{t} [\mathbf{U}(t,r)\cdot\Delta\mathbf{G}(r)]\cdot[\mathbf{U}(t,r)\cdot\Delta\mathbf{G}(r)]\, dr \\ &+ 2\int_{t_0}^{t}\int_{0+}^{r} [\mathbf{U}(t,r)\cdot\Delta\mathbf{G}(r)]\cdot[\mathbf{U}(t,r-s)\cdot\Delta\mathbf{G}(r-s)]\, ds\, dr. \end{aligned}$$

As before, note that the integral over $s$ can be suppressed by defining a new vector $\mathbf{H}[\mathbf{x}(r)]$ via

$$\begin{aligned} \int_{0+}^{r} \mathbf{U}(t,r-s)\cdot\Delta\mathbf{G}(r-s) ds &= \mathbf{U}(t,r)\cdot\int_{0+}^{r}\mathbf{U}(r,r-s)\cdot\Delta\mathbf{G}(r-s) ds \\ &= \mathbf{U}(t,r)\cdot\mathbf{H}[\mathbf{x}(r)]. \end{aligned} \tag{14}$$

Figure 2 schematically illustrate a piece of the trajectory $\mathbf{x}$. The true vector field, $\mathbf{G}$, is indicated by the arrows that are tangent to this trajectory. Because the dynamics of the model, Eq. (2), is close to the true dynamics $\mathbf{F}$ must point in a direction that is very close to the direction of $\mathbf{G}$. Thus $\Delta\mathbf{G} = \mathbf{F} - \mathbf{G}$ is essentially perpendicular to either $\mathbf{F}$ or $\mathbf{G}$. In addition we conjecture that the projection of $\Delta\mathbf{G}$ onto a $d-1$ dimensional hyperplane perpendicular to the flow takes on all possible orientations as one follows the flow along the attractor. Thus, although $\Delta\mathbf{G}$ always takes on the same direction at a particular location on the attractor, this direction will be different for different locations on the attractor. This ergodic assumption permits us to equate time averages with phase space averages and obtain

$$\begin{aligned} \langle I_1\rangle_T &= \int_{t_0}^{t}\left\langle |\mathbf{U}(t,r)\cdot\Delta\mathbf{G}(r)|^2\right\rangle dr + 2\int_{t_0}^{t}\langle[\mathbf{U}(t,r)\cdot\Delta\mathbf{G}(r)]\cdot[\mathbf{U}(t,r)\cdot\mathbf{H}(r)]\rangle\, dr \\ &= A^2, \end{aligned} \tag{15}$$

The form of the integral, $\langle I_1\rangle_T$, shown in Eq. (15) is exact. $A$ is a nontrivial function of modeling errors, but is *not* a function of noise.

A more complete analysis of the integrals in Eqs. (12) and (15) can be found in Appendices B and C. For our present purpose the forms of Eqs. (12) and (15) are sufficient. The analysis of Eq. (9) has shown that the deviation level, as defined by Eq. (8), can be approximated by the following function of $\sigma$ and $\Delta\mathbf{G}$

$$\left[\langle|\mathbf{z}|^2\rangle_T\right]^{1/2} = \left[A^2 + (\sigma B)^2\right]^{1/2}, \tag{16}$$

where $B^2$ and $A^2$ are given by Eqs. (12) and (15), respectively. The first question we asked in Section I concerned the effect of additive noise on the synchronization deviation level. Equations (12), (15) and (16) provide a theoretical answer to that question.

The second question asked in Section I concerned modeling errors and the effects of drift in the dynamics of the driving signal. Both of these effects will influence $A$ but neither will influence $B$. The final portion of this section derives an explicit expression for how these effects influence $A$. We begin by inserting Eq. (5) into Eq. (14). The result is

$$\begin{aligned} \mathbf{H}[\mathbf{x}(r)] &\simeq \int_{0+}^{r}\mathbf{U}(r,r-s)\cdot\Delta\mathbf{G}'(r-s)\, ds + \left[\frac{\partial}{\partial\mathbf{p}}\int_{0+}^{r}\mathbf{U}(r,r-s)\cdot\Delta\mathbf{G}'(r-s)\, ds\right]\cdot\delta\mathbf{p} \\ &= \mathbf{H}'[\mathbf{x}(r)] + \left[\frac{\partial}{\partial\mathbf{p}}\mathbf{H}'[\mathbf{x}(r)]\right]\cdot\delta\mathbf{p} \end{aligned}$$

where the second equation serves to define $\mathbf{H}'$ and $\partial\mathbf{H}'/\partial\mathbf{p}$ in an obvious manner. Inserting this expression for $\mathbf{H}$, as well as Eq. (5), into Eq. (15) results in

$$A^2 \simeq A'^2 + \left[\frac{\partial}{\partial\mathbf{p}}\int_{t_0}^{t}\left\langle|\mathbf{U}(t,r)\cdot(\Delta\mathbf{G}'(r)+\mathbf{H}'(r))\,|^2\right\rangle\, dr\right]\cdot\delta\mathbf{p}, \tag{17}$$

where $A'^2$ is defined by

$$A'^2 = \int_{t_0}^{t} \left\langle \left|\mathbf{U}(t,r)\cdot\Delta\mathbf{G}'(r)\right|^2 \right\rangle dr + \int_{t_0}^{t} \left\langle [\mathbf{U}(t,r)\cdot\Delta\mathbf{G}'(r)]\cdot[\mathbf{U}(t,r)\cdot\mathbf{H}'(r)] \right\rangle dr.$$

Equation (17) says that, for fixed noise size and type, the deviation level, as given by Eq. (16) will rise linearly with changes in the parameters of the driving dynamics. By obtaining numerical values for the two terms in Eq. (17) we will be able to predict the rise in the deviation level when the dynamics of the driving signal is drifting. The ability to make this prediction is the center of the non-destructive testing application we discuss in the conclusion.

Equations (12), (15), (16), and (17) (and the material contained in the appendices) represent the principle theoretical results of this paper. In the next section we perform simple numerical experiments in an attempt to test the behavior predicted by these equations.

## III. NUMERICAL EXPERIMENTS

In this section of our paper we present the results of the numerical experiments we have performed. In Section II we addressed changes in the synchronization deviation level as a function of both noise level, $\sigma$, and $\Delta\mathbf{G}$. Our numerical experiments examine each of these issues separately. In addition all of the data sets used for our numerical experiments were recorded from physical experiments. Some of our numerical experiments are performed on data sets taken from an electronic circuit while others are performed on data sets taken from a mechanical system (a vibrating wire). All of our numerical experiments we used the third component of the embedded time series as the driving term. Thus, $E_{33} = \epsilon \neq 0$ in Eq. (3)

A block diagram for the electronic circuit whose behavior we investigate is shown in Fig. 3. It consists of a nonlinear amplifier, $N$, which transforms input voltage, $x(t)$, into output, $\alpha f(x)$. The parameter $\alpha$ characterizes the gain of $N$ around $x = 0$. The nonlinear amplifier has linear feedback which contains a series connection to a low-pass filter ($RC'$) and $LC$ resonance. More detailed discussions of this circuit can be found in Refs. [13,18,25,27].

A scalar time series, $s(n\Delta t) = s(n)$, (where $\Delta t$ is the sampling interval) is experimentally measured for different values of $\alpha$. For modeling purposes the sampling interval is assigned the numerical value $\Delta t = 0.02$. The $\alpha$ values we initially investigated are $\alpha = 17.4$, which we call DX1, and $\alpha = 18.9$, which we call DX2. We report the following means, $\langle s\rangle = 3.83$ or 0.55, and standard deviations, $\sigma_s = 2.53$ or 4.53 for DX1 and DX2, respectively. The time delay method

$$\mathbf{x}(n) = [s(n),\, s(n+T), \ldots, s(n+(d-1)T)]$$

is used to reconstruct the working phase space of the dynamics. It was previously determined that the optimal time delay and embedding dimension are $T = 10$ and $d = 3$, respectively [18]. The attractors obtained by embedding 5000 data vectors from DX1 and DX2 are shown in Figs. 4a and b, respectively.

When $\alpha = 17.4$ the dynamics lives on one of two disjoint attractors. One of these attractors is shown in Fig. 4a. The other attractor is related to this one by an inversion symmetry and is not shown. A time series would measure one or the other of these attractors but not both. As $\alpha$ increases these two attractors eventually merge at $\alpha = \alpha_c$ in what has been called a crisis, or a symmetry increasing bifurcation [28,29]. The attractor associated with the DX2 time series arises after the merge ($18.9 > \alpha_c$).

A second experimental scalar time series is obtained from an apparatus used to study nonlinear vibrations in a wire. Briefly, the apparatus consists of a mount holding a tensioned wire through which an alternating current is passed. The frequency of the current is near the fundamental of the free oscillation frequency of the wire. This current excites forced vibrations when the wire is placed in a permanent magnetic field. As the current's amplitude and frequency are varied the system can undergo a torus-doubling route to chaos [30]. Optical detectors are used to measure the transverse amplitude displacement of the wire. The transverse displacements of the wire form the time series that is analyzed.

The apparatus used here is a significantly improved version of that reported in [31]. Rare earth ceramic magnets are used to provide the permanent magnetic field and custom digitizing and controlling circuits were build around a digital signal processor [32]. Typical experimental parameters for this wire are: length (0.07 m), mass per unit length ($3.39 \times 10^{-4}$ kg/m), density ($2.1 \times 10^{4}$ kg/m$^3$), Young's modulus ($197778 \times 10^{6}$ N/m$^2$), and magnetic field strength (0.2 Tesla).

The chaotic time series analyzed here is obtained at a forcing frequency of 1.384 kHz. The chaotic response of the wire occurs in the slowly varying modulations of the envelope of the oscillations and has a characteristic frequency of about 9 Hz in this particular experiment.

The data set consists of 128,000 scalar 16-bit measurements of the vertical transverse wire displacement. The signal has been cleaned using a technique developed by Schreiber [33,34] (after cleaning noise levels $< 1$ %) and has an RMS response amplitude drift of $\sim 1$ % over the entire digitized record (about 200 seconds). The signal is over digitized, about 150 points were obtained per chaotic cycle. Typically only a fraction of this data set is analyzed.

Because the values of the records in the wire data set are typically $\sim 10^4$ we rescale this data to have zero mean and standard deviation $\sigma_s = 4.0$. The time delay method is used to embed the rescaled data. The average mutual information and false near neighbor methods indicated that the optimal time delay and embedding dimension are $T = 39$ and $d = 3$, respectively [23,35,36]. For modeling purposes the sampling interval is assigned the numerical value $\Delta t = 0.05$. The attractor obtained by embedding 10,000 data vectors from the vibrating wire is shown in Fig. 5.

Having reconstructed the attractors in a working phase space we constructed model vector fields for the dynamics on these attractors. The models were global ODE in the form of Eq. (2), and were constructed from the time series shown in Figs. 4 and 5.

In our first set of numerical experiments we examined the behavior of $[\langle|\mathbf{z}|^2\rangle_T]^{1/2}$ as a function of $\sigma$. For each numerical experiment on DX1 we used the same time series, $\mathbf{x}$, as our clean driving signal. The same is true for DX2 and the vibrating wire. (Care was taken to insure that the series used as our clean driving signals were not the ones used to construct the model vector fields.) To obtain a noisy driving signal we numerically generated a noise signal, $\hat{\mathbf{u}}$, ($\hat{\mathbf{u}}$ is constructed to have unit size) and added it to $\mathbf{x}$ to form $\mathbf{x} + \sigma\hat{\mathbf{u}}$.

We employed two different methods to generate the noise. The first method generated a Gaussian noise signal via a random number generator. The time series of points in the noise signal are chosen to have mean zero and standard deviation one. The second type of noise signal was generated by Fourier transforming DX1 and then randomizing the phases of the Fourier coefficients. After inverting the randomized Fourier transform the time series was rescaled to have mean zero and standard deviation one. Thus, our second procedure produces a noise signal that has the same power spectrum and autocorrelation function as the experimentally recorded time series, DX1. We call this inband noise and argue that this type of noise would be very difficult to detect, against the background of the chaotic driving signal, using standard Fourier methods. Our use of inband noise is similar, in spirit, to the use of "surrogate data" to test for determinism in time series data [37,38]. We use the same procedures to generate inband noise signals from DX2, and the vibrating wire.

As a pictorial example of these two types of noise Figs. 6 show the attractors constructed from noisy DX2 signals. In Fig. 6a a time series from DX2 has been contaminated with Gaussian noise of size $\sigma = 1$. In Fig. 6b the same time series has been contaminated with inband noise of size $\sigma = 1$. We remark that inband noise gives the illusion of being dynamical (it looks like a clean signal embedded in too low an embedding dimension) [39,40].

Our numerical tests of the effects of noise on synchronization involved the following steps. First, we selected a value for $\epsilon$, the coupling between the driving and the response system, somewhat above the threshold level for synchronization [18]. During this first set of numerical experiments the noise level will rise. Therefore, we do not want $\epsilon$ too close to the critical value $\epsilon_c$. We then synchronized the model to the noisy driving signal, $(\mathbf{x} + \sigma\hat{\mathbf{u}})$, via Eq. (3) and recorded $|\mathbf{z}|^2 = |\mathbf{y} - \mathbf{x}|^2$, (the lower curve in Fig. 1). We always used the model for DX1 when $\mathbf{x}$ was a DX1 time series, the model for DX2 when $\mathbf{x}$ was a DX2 time series, and the model for the vibrating wire when $\mathbf{x}$ was a vibrating wire time series. Likewise, when using inband noise we used inband noise from the appropriate system.

For DX1 and DX2 we recorded 5000 time steps after discarding the first 500 as a transient. For the vibrating wire we recorded 10,000 time steps after discarding the first 5000 time steps as a transient. From the recorded values of $|\mathbf{z}|^2$ a time average, $\langle|\mathbf{z}|^2\rangle_T$, was calculated. This process was repeated for a variety of $\sigma$ values covering three decades of magnitude and both types of noise. Finally, we increased $\epsilon$ to a value well above the threshold for synchronization and repeated all test on all types of noise.

As indicated above, if $\epsilon < \epsilon_c$ then the coupling will not be strong enough to synchronize $\mathbf{F}$ to the time series. We have found, from numerical testing on raw data (data without added noise) that the synchronization threshold for DX1, as well as DX2, is $\epsilon_c \simeq 2$. For our numerical experiments with additive noise $\epsilon = 5$ was the small value and $\epsilon = 20$ was the large value. For the vibrating wire the threshold is $\epsilon_c \simeq 0.5$, and we tested at $\epsilon = 0.75$ and $\epsilon = 3$.

In Section II we have argued that the following scaling law holds for the deviation level

$$[\langle|\mathbf{z}|^2\rangle_T]^{1/2} = [A^2 + (\sigma B)^2]^{1/2}$$

where $B^2$ and $A^2$ are the simple integrals indicated in Eqs. (12) and (15), respectively. As it stands this equation is not completely informative. Before concluding that the model is, or is not, synchronized to the time series one should normalize $\langle|\mathbf{z}|^2\rangle_T$ by the level of the top curve shown in Fig. 1. (By definition, the deviation level of the top curve is associated with trajectories that are not synchronized. Hence, it represents the most logical choice for normalization.) If the time average of the top curve is $D^2$ then the quantity that determines the normalized synchronization deviation level is $[\langle|\mathbf{z}|^2\rangle_T]^{1/2}/D$.

The results of our numerical experiments are shown in Figs. 7–9 where we have plotted the normalized synchronization deviation levels, $[\langle|\mathbf{z}|^2\rangle_T]^{1/2}/D$ as functions of $\sigma$. As predicted the normalized deviation levels rise as $\sigma$ increases. In order to compare Eq. (16) to the results shown in Figs. 7–9 we need to determine numerical values for $A/D$ and $B/D$. We have used the numerical values of $[\langle|\mathbf{z}|^2\rangle_T]^{1/2}$ and $\sigma$ along with a least squares algorithm to fit Eq. (16) and thereby obtain values of $A/D$ and $B/D$. The results of this procedure yield the curves shown in the figures. It is clear from the figures that the scaling law derived in Section II is an excellent match to the data obtained from the numerical experiments.

We noticed some unusual behavior, which we call bursting, when the noise level, $\sigma$, was large. An example of a burst for inband noise, $\sigma = 1$, $\epsilon = 5$, and the DX1 model is shown in Fig. 10. The figure indicates that for a small number of steps the noise has caused $\mathbf{F}$ to loose synchronization with $\mathbf{x}$. (Similar bursts have been reported by Mossayebi et. al [11].) This loss of synchronization occurs very infrequently and only for large noise levels. In fact we only found bursting for DX1 when inband noise was used. The noise levels and coupling strengths were $\sigma > 0.45$ when $\epsilon = 5$, and $\sigma = 1.0$ when $\epsilon = 20$. These values of $\sigma$ correspond to noise levels of approximately 20 % and 40 % respectively (percent noise is identified as 100 $(\sigma/\sigma_s)$). Furthermore, for the indicated values of $\sigma$ and $\epsilon$ we only found one burst for the run of 5000 time steps. For weak coupling the burst was just under 200 time steps in length, while for large coupling the burst was just under 50 time steps in length.

For DX2 we did not observe bursting. For the vibrating wire we noticed bursting for large values of the noise level. These bursts occurred for all cases when $\sigma = 1$, which corresponds to noise levels of $\sim$ 25 %. They also occurred when the coupling was weak ($\epsilon = 0.75$) and $\sigma > 0.45$. Since the bursts were very rare and relatively short we have not included there values of $|\mathbf{z}|^2$ when performing the time averages indicated by Eq. (8).

The appearance of bursts in our experiments may be connected with on-off intermittency provoked by the presence of noise. On-off intermittency has recently been studied by Platt et. al. [41] as well as Ott and Sommerer [42] and is associated with the stability properties of multi-dimensional integral manifolds which may contain chaotic attracting sets (we found these papers to be particularly convincing [43]). The latter paper discusses on-off intermittency in the noise free case and argues that bursts are typical behavior which accompany on-off intermittency. One portion of the paper discusses this behavior in the context of synchronization and implies that bursts would be observed if synchronization looses stability and experiences what they call a nonhysteretic bifurcation of the invariant manifold.

The difference between our system and that discussed by Ott and Sommerer is that our noise free system is stable and well beyond the threshold of stability. However, as suggested by Platt et. al and others [44] a high noise level could result in dynamics where the coupling parameter, $\epsilon$ is *effectively* below the synchronization threshold, $\epsilon_c$, for short periods of time. If this is occurring then the bursts we see could be the result of what one would call a noise induced nonhysteretic bifurcation.

The final set of numerical experiments we performed involved determining the behavior of the deviation level, $[\langle|\mathbf{z}|^2\rangle_T]^{1/2}$, as a function of changes in the dynamics of the driving signal, $\Delta\mathbf{G}$. In order to perform our numerical tests we recorded six time series from the circuit corresponding to $\alpha = 17.10$, 17.23, 17.36, 17.49, 17.62, and 17.75. These time series were measured with a different value of $R$, (see Fig. 3) than the previous time series. Since the parameters of the circuit have changed the exact dynamics of these signals will not match either DX1 or DX2. However, the structure of the attractors is qualitatively similar to the attractors associated with DX1 (see Fig. 4a).

A model, Eq. (2), was constructed from the data associated with $\alpha = 17.75$. The model was then subjected to driving from time series associated with $\alpha = 17.75, \ldots, 17.10$. For each value of $\alpha$ the deviation level was measured. The results of this experiment are shown in Fig. 11. The solid lines in Fig. 11 represent lines of best fit through the data. The solid circles, squares, diamonds, and triangles are associated with coupling strengths of $\epsilon = 1$, 2, 3, and 4, respectively. Clearly, all of the solid symbols indicate a linear rise in the deviation level with respect to changes in $\Delta\alpha$. For comparison purposes the open circles represent a test case with a coupling strength slightly smaller than $\epsilon_c$.

We expect the deviation level to rise as $\alpha$ deviates from $\alpha = 17.75$. This rise is assumed to be due, predominantly, to changes in the dynamics of the driving signal. In order to verify this assumption the noise levels in the driving signal should be small. We believe that this is the case for two reasons. First, the modeling procedure allows one to make a rough guess at the noise level in the signal [18]. We find that when modeling signals from our circuit the noise levels are $\sim 10^{-3}$. Second, Figs. 7 and 8 indicate that for noise levels below $\sim 10^{-2}$ the level of synchronization is independent of $\sigma$. These two facts lead us to believe that the level of synchronization in Fig. 11 is being dominated by the $A$ term in Eq. (16). Therefore, changes in the level of synchronization in our numerical experiments should be associated with changes in the driving signal, $\Delta\mathbf{G}$.

If the amplitude of the noise is small then Eqs. (15), (16), and (17) show that

$$\log\left([\langle|\mathbf{z}|^2\rangle]^{1/2}\right) \simeq \log(A') + \frac{1}{2}\left(\frac{S}{A'}\right)^2 |\Delta\alpha|,$$

for suitably defined $S(\epsilon)$ (see Eq. (17)). This equation indicates that the deviation level should rise linearly with $|\Delta\alpha|$ which is exactly what is shown in Fig. 11. For the numerical experiments $|\Delta\alpha| = 0.13$ which represents a change of approximately 1 % in the value of $\alpha$. Therefore, the assumption that $|\Delta\alpha|$ is small is valid for the numerical experiments we have performed.

## IV. SUMMARY AND CONCLUSIONS

In this paper we investigated the behavior of chaotic synchronization in the presence of two common problems. The first problem is associated with additive noise in the data. This problem occurs whenever real experimental data is used as the driving signal. The second problem is associated with differences between the dynamics of the driving system and the dynamics of the response system. The second problem is also quite common but comes in many disguises, some of which we will discuss later.

The numerical experiments involved constructing models of the dynamics of the driving system from experimentally measured time series. These models were then driven by noisy time series from the systems they were designed to model. The data used for the numerical experiments came from two different experimental systems. The first system is a nonlinear circuit and the second is a vibrating wire. In both cases the raw data was in the form of scalar time series.

In an attempt to provide quantitative results we defined a measure of the deviation from synchronization between the clean driving signal and the driven model. If the clean driving signal is denoted by $\mathbf{x}$ and the dynamics of the driven model is denoted by $\mathbf{y}$ then the synchronization deviation level is given by the following time average $\left[\langle|\mathbf{z}|^2\rangle_T\right]^{1/2}$, where $\mathbf{z} = \mathbf{y} - \mathbf{x}$ (see Eq. (8) in Section II).

To study the effect of additive noise in the driving signal on the synchronization deviation level we added Gaussian or inband noise with standard deviation $\sigma$ to the experimental signal to produce a driving signal, $(\mathbf{x} + \sigma\hat{\mathbf{u}})$. Models were then driven by these signals for various values of $\sigma$ and the synchronization deviation levels measured. Theoretical analysis in Section II indicates that $\left[\langle|\mathbf{z}|^2\rangle_T\right]^{1/2} = \left[A^2 + (\sigma B)^2\right]^{1/2}$ where $B$ and $A$ are given by Eqs. (12) and (15). The analysis indicates that $A$ is a function of errors in our modeling of the driving dynamics while $B$ is a function of the autocorrelation statistics of the noise. Both $A$ and $B$ are independent of $\sigma$, the magnitude of the noise. The results of our numerical experiments are shown in Figs. 7–9 and confirmed this scaling law. They also indicated that synchronization is very robust to noise in the driving signal. The figures show that synchronization persists even with noise levels as high as 40 %.

To study the effects of drift in the dynamics of the driving signal we used a time series from the circuit to build a model of the dynamics of the circuit. We then obtained time series measurements from the circuit after making small changes in $\alpha$, a parameter that influences the nonlinearity of the circuit. The model was then subjected to driving from time series associated with the various values of $\alpha$. For each case the synchronization deviation level was recorded. Theoretical analysis in Sections II and III indicates that $\left[\langle|\mathbf{z}|^2\rangle_T\right]^{1/2}$ will rise linearly with $\Delta\alpha$. The results of our numerical experiments are shown in Fig. 11 and confirm this scaling law.

The quantities $B$ and $A$ are functions of the coupling strength $\epsilon$. In the appendices we analyze the functional dependence of $B$ and $A$ on $\epsilon$. We derive a simple expression for $B$ (see Eq. (B5)) for delta correlated noise (for example, Gaussian noise). Finally, by knowing $A$ we can obtain an order of magnitude approximation for $\left[\langle|\Delta\mathbf{G}|^2\rangle\right]^{1/2}$, the average error in our model of the vector field (see Appendix B).

We close this section with a discussion of the relevance of these results to a variety of possible applications. To date, the primary suggested application for synchronization involves communications [27,45–50]. All of the proposed communication methods that use a chaotic carrier to mask the signal of interest require synchronization (identical synchronous motion) between the transmitter generating the chaotic carrier and the receiver. Clearly, any noise in the transmission channel will introduce deviations from synchronization between the transmitter and the receiver. These deviations are not part of the signal of interest and affect the quality of the recovered signal. Furthermore, if the transmission channel distorts the signal (possibly because of some limit in its transmission capabilities) then the dynamics of the received signal will differ from the dynamics of the transmitted signal. This change in the dynamics of the transmitted signal will affect our ability to synchronize the receiver to the transmission. Therefore, one must understand the effect of noise and changes in the driving dynamics on synchronization if synchronization is to be used as a communications method.

Another possible application of synchronization is system identification for chaotic sources. Suppose that the only thing we know about a particular system is a previously measured clean time series, $\mathbf{x}$. At some later time a second time series, $\mathbf{x}'$, is received and we wish to determine whether or not $\mathbf{x}$ and $\mathbf{x}'$ were produced by the same system. To answer this question one might construct a model of the dynamics that produced $\mathbf{x}$ and attempt to synchronize this model to $\mathbf{x}'$. From the stand point of synchronization we say that if the model synchronizes to $\mathbf{x}'$ then $\mathbf{x}$ and $\mathbf{x}'$ have the same source, and we have identified the signal. Clearly, the noise level in $\mathbf{x}'$ will affect our ability synchronize the model to $\mathbf{x}'$. Furthermore, since we only have a time series with which to construct the model it is certain that the dynamics of the model will not be *exactly* the same as the dynamics that produced $\mathbf{x}$. This modeling error will affect our ability to synchronize the model to $\mathbf{x}'$. Therefore, one must understand the effect of noise and modeling error on synchronization if synchronization is to be used as a system identification method.

The final application of synchronization we discuss is non-destructive testing. Consider some new device which is to be placed into the field as part of its normal operational life. Prior to placement the device is driven by a calibrated external driving source and a time series is recorded. Once this "shake down" test is completed the device is placed into the field, a model of the dynamics is constructed from part of the time series, and the synchronization deviation level between the model and the rest of the time series is recorded. After some time we wish to determine whether or not the device is in need of maintenance. To accomplish this we drive the device by a calibrated external driving source, and record a new time series. Next, we attempt to synchronize this new time series to the previously constructed model. Since the device has experienced some ware its dynamics will have changed and the synchronization deviation level will have risen. By monitoring this rise one can determine whether or not the device is in need of maintenance. Therefore, one must understand the effect of drift in the dynamics of the driving system on synchronization if synchronization is to be used as a non-destructive testing method.

## V. ACKNOWLEDGMENTS

The authors would like to thank Drs. Lev Tsimring and Sid Sidorowich for graciously providing advice while discussing various aspects of this research. We also thank Dr. Tim Molteno of Otago University for his assistance in providing the wire data set. N. Rulkov was supported by grant number DE-FG03-90ER14148 from the Department of Energy. N. Tufillaro was supported by the Department of Energy.

## APPENDIX A: THE $\epsilon \to \infty$ LIMIT

In this short appendix we briefly re-examine the analysis presented in Section II in the limit of $\epsilon \to \infty$. We are primarily interested in whether or not $B$ and $A$ exist and are well defined in this limit. Our analysis begins with Eq. (4) which we rewrite at this time

$$\frac{d\mathbf{z}}{dt} = [\mathbf{DF}(\mathbf{x}) - \mathbf{E}] \cdot \mathbf{z} + \sigma \mathbf{E} \cdot \hat{\mathbf{u}} + \Delta \mathbf{G}(\mathbf{x}).$$

Assume, without loss of generality, that the nonzero component of $\mathbf{E}$ is $E_{dd} = \epsilon$. Then the $d$th equation can be rewritten as

$$\frac{1}{\epsilon}\frac{dz_d}{dt} = \frac{1}{\epsilon}\frac{\partial F_d(\mathbf{x})}{\partial \mathbf{x}} \cdot \mathbf{z} - [z_d - \sigma\eta] + \frac{1}{\epsilon}\left[F_d(\mathbf{x}) - G_d(\mathbf{x})\right].$$

Letting $\epsilon \longrightarrow \infty$ this equation reduces to $z_d = \sigma\eta$ [18]. In this limit the linearized equation of motion for $\mathbf{z}$ becomes the following $d-1$ dimensional equation of motion

$$\frac{d\mathbf{z}'}{dt} = \mathbf{DF}'(\mathbf{x}) \cdot \mathbf{z}' + \sigma \hat{\mathbf{u}}' + \Delta \mathbf{G}'(\mathbf{x}), \tag{A1}$$

where $\mathbf{z}'$ and $\Delta\mathbf{G}'$ are vectors composed of the first $d-1$ components of $\mathbf{z}$ and $\Delta\mathbf{G}$, respectively. The matrix $\mathbf{DF}'$ is a $d-1$ by $d-1$ matrix obtained by eliminating the $d$th row and column from $\mathbf{DF}$, and $\hat{\mathbf{u}}'$ is the following $d-1$ dimensional vector

$$u'_\beta = \eta \frac{\partial F_\beta(\mathbf{x})}{\partial x_d}.$$

Notice that all reference to $\epsilon$ has been eliminated from Eq. (A1) and each term is well defined, although the noise term is somewhat unusual since $\hat{\mathbf{u}}'$ depends on $\mathbf{x}$ as well as time. The dynamics of Eq. (A1) evolves in a $d-1$ dimensional phase space. This type of equation is what one would expect to derive from the Pecora and Carroll form of synchronization. In the absence of noise and modeling errors the Lyapunov exponents of Eq. (A1) arise from $\mathbf{DF}'$ and are usually called the conditional Lyapunov exponents [2]. The solution to Eq. (A1) can be obtained by applying the same techniques used to solve Eq. (4). The result is

$$\mathbf{z}'(t) = \mathbf{U}'(t, t_0) \cdot \mathbf{z}'(t_0) + \int_{t_0}^{t} \mathbf{U}'(t, r) \cdot [\Delta\mathbf{G}'(r) + \sigma\hat{\mathbf{u}}'(r)]\, dr,$$

where $\mathbf{U}'(t, t_0)$ is defined as the operator that evolves $\mathbf{z}'(t_0)$ into $\mathbf{z}'(t)$ in the absence of noise and modeling errors. It can be formally written as

$$\mathbf{U}'(t,t_0) = \exp\left[\int_{t_0}^{t} \mathbf{DF}'(r)dr\right].$$

We end this appendix by pointing out that the synchronization deviation level defined by Eq. (8) can be redefined for the $d-1$ dimensional vector $\mathbf{z}'$ in a straightforward manner. An equation similar to Eq. (9) can be obtained by forming $|\mathbf{z}'(t)|^2$ and terms similar to $B$ and $A$ can be defined. We will not derive explicit expressions for $B$ and $A$ because we have accomplished our purpose by demonstrating that they exist and are well defined in the $\epsilon \longrightarrow \infty$ limit.

## APPENDIX B: DETAILED ANALYSIS OF $B$ AND $A$

In Section II we derived a theoretical prediction for the behavior of the synchronization deviation level, Eq. (16), as a function of the noise level, $\sigma$. This dependence was then verified in Section III. The analysis involved two terms which we called $B$ and $A$ and are defined by Eqs. (12) and (15), respectively. In this appendix we present a partial analysis of the dependence of $B$ and $A$ on the coupling strength, $\epsilon$. The results in this appendix are somewhat speculative. We have presented them for two reasons: (1) the dependence of $B$ and $A$ on $\epsilon$ has direct impact on the central issue of this paper (the behavior of the synchronization deviation level). (2) our analysis is the first attempt (that we are aware of) to analytically study the effect of $\epsilon$ on the synchronization deviation level for the type of coupling we have used.

We begin the analysis with $B$ which is defined by Eq. (12) which we rewrite in the form

$$B^2 = B_0^2[1+2R_B]$$

were $B_0$ and $R_B$ are defined by

$$B_0^2 = \epsilon^2 k(0)\int_{t_0}^{t}\left\langle \mathbf{V}\cdot\left[\mathbf{U}^T(t,r)\cdot\mathbf{U}(t,r)\right]\cdot\mathbf{V}\right\rangle dr \tag{B1}$$

and

$$R_B = \frac{\epsilon^2 k(0)}{B_0^2}\int_{t_0}^{t}\left\langle \mathbf{V}\cdot\left[\mathbf{U}^T(t,r)\cdot\mathbf{U}(t,r)\right]\cdot\mathbf{B}(r)\right\rangle dr. \tag{B2}$$

where the superscript $T$ indicates the transpose.

An important special case is delta correlated noise where $k(s) = k(0)\,\delta(s)$. (For example, Gaussian noise and uniform noise are delta correlated.) When the noise is delta correlated it is obvious from Eq. (11) that $\mathbf{B} = \mathbf{0}$ which leads to $B^2 = B_0^2$. The behavior of $\mathbf{B}$, and hence $B^2$, for other situations is a much more complicated matter. It depends on $k(s)$, the autocorrelation function of the assumed stationary noise. In light of the importance of delta correlated noise we will address $B_0^2$ in detail.

The matrix $\left[\mathbf{U}^T(t,r)\cdot\mathbf{U}(t,r)\right]$ is symmetric, hence its eigenvectors are orthonormal. For convenience define $\tau$ by $\tau = t-r$, and let the eigenvectors and eigenvalues of this matrix be denoted, respectively, by $\hat{\mathbf{e}}^{(\beta)}(r,\tau)$, and $\exp[2\tau\lambda_\beta(r,\tau)]$, for $\beta = 1,\ldots,d$. We recognize $\lambda_\beta(r,\tau)$ as the $\beta$th local Lyapunov exponent [23,51]. The local Lyapunov exponents control the local expansions and contractions for the trajectory that begins at $\mathbf{x}(r)$ and ends at $\mathbf{x}(t)$. Expand $\mathbf{V}$ in terms of the eigenvectors of $\left[\mathbf{U}^T(t,r)\cdot\mathbf{U}(t,r)\right]$ to get

$$\mathbf{V} = \sum_{\beta=1}^{d} v^{(\beta)}(r,\tau)\,\hat{\mathbf{e}}^{(\beta)}(r,\tau).$$

Inserting this expansion and the eigenvalues of $\left[\mathbf{U}^T(t,r)\cdot\mathbf{U}(t,r)\right]$ into Eq. (B1), and using the orthogonality of the eigenvectors results in

$$B_0^2 = \epsilon^2 k(0)\sum_{\beta=1}^{d}\int_{0}^{t-t_0}\left\langle\left[v^{(\beta)}(r,\tau)\right]^2\exp\left[2\tau\lambda_\beta(r,\tau)\right]\right\rangle d\tau. \tag{B3}$$

Equation (B3) is exact. Taking the ensemble average indicated in Eq. (B3) will result in the loss of information about the individual locations, $\mathbf{x}(r)$. Thus, the ensemble average shown in Eq. (B3) can only be a function of $\tau$. In order to perform the indicated ensemble average we need to know the joint probability distribution of the local

Lyapunov exponents, $\lambda_\beta(r,\tau)$, and the orientations of the eigenvectors, $\hat{\mathbf{e}}^{(\beta)}(r,\tau)$. This probability distribution is unknown, and we know of no papers that have attempted to study this joint probability distribution. In order to proceed with our calculation we will make three assumptions/approximations regarding this probability distribution. Before explicitly stating these assumptions/approximations it is useful to discuss the probability distribution of the local Lyapunov exponents.

For each value of $\tau$ there are $d$ probability distributions of the local Lyapunov exponents, one for each $\lambda_\beta(r,\tau)$. Denote these probability distributions by $\rho_\beta(\lambda)$, where the dependence on $\tau$ is understood. Figure 12 shows normalized histograms of $\rho_\beta(\lambda)$ for $\beta = 1$, 2, and 3. The histograms are obtained by calculating the local Lyapunov exponents for the model associated with data set DX1 (see Section III). The evolution time is $\tau = \Delta t = 0.02$, we used 10,000 different initial conditions, and the histograms are normalized so that $\int_{-\infty}^{\infty} \rho_\beta(\lambda) d\lambda = 1$.

We have examine $\rho_\beta(\lambda)$ for $\beta = 1$, 2, and 3 and $\tau = \Delta t, 2\Delta t, \ldots, 100\Delta t$. As $\tau$ increases the location of the center of mass of $\rho_1(\lambda)$ decreases and eventually becomes negative. This is to be expected since negative global Lyapunov exponents for the driven equation of motion, Eq. (3), are required for the stability of synchronization [1,2,7–10]. What is not obvious, and is clear from Fig. 12, is that for small values of $\tau$ it is possible for *all* of the local value of $\lambda_1(r,\tau)$ to be positive. This leads to the interesting conclusion that for this type of coupling *all* locations on the attractor have at least one expanding direction. Thus, nearby trajectories are always diverging in this eigendirection and it is only the rotation of the eigendirections, as the trajectory moves along the attractor, that ultimately causes synchronization. Behavior that is similar to the type we have just discussed is observed for all of the test cases discussed in Section III. For this reason we believe that our results are generic to low dimensional dynamical systems that are driven in the manner of Eq. (3).

The histograms of $\rho_\beta(\lambda)$ for values of $\tau \neq \Delta t$ usually look similar to the ones shown in Fig. 12, namely sharply peaked. However, exceptions, where the histograms are bimodal or relatively flat do occur. These observations imply that for small values of $\tau$ the $\rho_\beta(\lambda)$'s are, typically, not Gaussian, or any other simple probability distribution.

Having partially examined the probability distributions of the local Lyapunov exponents we are ready to make some approximations regarding the joint probability distribution of the $\lambda_\beta(r,\tau)$'s and the $\hat{\mathbf{e}}^{(\beta)}(r,\tau)$'s.

1. The first assumption to be addressed involves whether or not the joint probability distribution will factor. From the definition of global Lyapunov exponents it follows that in the large $\tau$ limit, the orientations of the eigenvectors are independent of the values of the local Lyapunov exponents. Hence, in this limit the joint probability distribution factors. In the small $\tau$ limit the orientations of the eigenvectors are not independent of the local Lyapunov exponents. However, it is possible that independence is a good approximation.

   Numerical experiments have indicated that the bulk of the $\lambda_\beta$'s occur within a narrow range of values corresponding to the peaks of $\rho_\beta(\lambda)$. Thus, if we choose local Lyapunov exponents at random from their natural distribution on the attractor then most of the chosen values will lie within this range which is associated with a large portion of the attractor.

   Furthermore, for small values of $\tau$ the $\exp[2\tau\lambda_\beta(r,\tau)]$ term will not vary much across the distributions of the local Lyapunov exponents. Hence, each $v^{(\beta)}(r,\tau)$ in the ensemble average will be multiplied by almost the same number [52].

   For these reasons we adopt the approximation of independence between the orientations of the eigenvectors, $\hat{\mathbf{e}}^{(\beta)}(r,\tau)$, and the values of the local Lyapunov exponents, $\lambda_\beta(r,\tau)$. After factoring the joint probability distribution the ensemble average in Eq. (B3) becomes

$$\left\langle \left[v^{(\beta)}(r,\tau)\right]^2 \exp\left[2\tau\lambda_\beta(r,\tau)\right] \right\rangle = \left\langle \left[v^{(\beta)}(r,\tau)\right]^2 \right\rangle \left\langle \exp\left[2\tau\lambda_\beta(r,\tau)\right] \right\rangle .$$

2. The second assumption we will address involves the probability distributions for the orientations of the eigenvectors for a fixed value of $\tau$. There are $d$ probability distributions, one for each value of $\beta$. A paper by Green and Kim is the only one we know of that has examined these orientations (it does so in the context of the Lorenz equations and the infinite $\tau$ limit) [53].

   To summarize the results of Green and Kim consider a particular point on the Lorenz attractor and imagine a plane that approximately spans the attractor at that point. Green and Kim found that the eigenvectors associated with the positive and zero Lyapunov exponents are in the plane while the eigenvector associated with the negative Lyapunov exponent is perpendicular to the plane. The orientation of the eigenvectors within the plane are controlled by the local changes in the velocity of the trajectory. At some locations on the attractor the positive direction is essentially orthogonal to the trajectory while at other locations on the attractor it is essentially parallel to the trajectory. We assume that these results hold for the systems we have studied.

When applied to our systems the results of Green and Kim indicate that, typically, the orientations of the eigenvectors will take on many values since the plane that locally spans the attractor has many different orientations. Thus, we assume that the probability distributions of the orientations of the $\hat{\mathbf{e}}^{(\beta)}(r,\tau)$'s will be quite broad and not drastically different for different values of $\beta$. This assumption implies that $\left\langle \left[v^{(\beta)}(r,\tau)\right]^2 \right\rangle \simeq \left\langle \left[v^{(\gamma)}(r,\tau)\right]^2 \right\rangle$ even when $\beta \neq \gamma$.

3. The final assumption we will address also involves the probability distributions of the orientations of the eigenvectors. In principle, as $\tau$ changes these distributions will change, and the ensemble averages, $\left\langle \left[v^{(\beta)}(r,\tau)\right]^2 \right\rangle$, will change. However, unless the $\tau$ dependence of the probability distributions is, in some sense, strong the ensemble averages will not change much. We know of no papers that have claimed to investigate the $\tau$ dependence of the probability distributions of the orientations of the $\hat{\mathbf{e}}^{(\beta)}(r,\tau)$'s. We will assume that the functional dependence is weak and that the ensemble averages, $\left\langle \left[v^{(\beta)}(r,\tau)\right]^2 \right\rangle$, are essentially independent of $\tau$.

These three assumptions/approximations, when taken together, allow us to rewrite Eq. (B3) as

$$B_0^2 \simeq \epsilon^2 k(0) \frac{1}{d} \sum_{\beta=1}^{d} \int_0^{t-t_0} \left\langle \exp\left[2\tau\lambda_\beta(r,\tau)\right]\right\rangle d\tau.$$

Earlier, we argued against the existence of a simple analytical expression for $\rho_\beta(\lambda)$. In this absence we choose to rewrite the ensemble average in this equation as

$$\begin{aligned} \left\langle \exp\left[2\tau\lambda_\beta(r,\tau)\right]\right\rangle &= \int_{-\infty}^{\infty} \rho_\beta(\lambda) \exp[2\tau\lambda_\beta(r,\tau)] d\lambda \\ &= \exp\left[2\tau\Lambda_\beta(\tau)\right], \end{aligned} \tag{B4}$$

where the last equality serves to define $\Lambda_\beta(\tau)$. With this definition $B_0$ is approximated by

$$B_0^2 \simeq \epsilon^2 k(0) \frac{1}{d} \sum_{\beta=1}^{d} \int_0^{\infty} \exp[2\tau\Lambda_\beta(\tau)] d\tau, \tag{B5}$$

where the upper limit of integration has been replaced by $\infty$. In Appendix C we present a detailed analysis of the integral in Eq. (B5), demonstrate that it does not diverge for the numerical examples investigated in Section III, and justify replacing the upper limit of integration by $\infty$.

A crude analysis that uses minor variations of assumptions/approximations 1, 2, and 3 can be performed on $R_B$. The result is

$$R_B \simeq \langle \mathbf{V} \cdot \mathbf{B} \rangle = \left\langle \int_0^{r-} \frac{k(s)}{k(0)} \mathbf{V} \cdot \left[\mathbf{U}(r,\tau) \cdot \mathbf{V}\right] d\tau \right\rangle \tag{B6}$$

We now turn our attention to $A$, given by Eq. (15), which we rewrite in the form

$$A^2 = A_0^2 [1 + 2R_A],$$

were $A_0$ and $R_A$ are defined by

$$A_0^2 = \int_{t_0}^{t} \left\langle \Delta\mathbf{G}(r) \cdot \left[\mathbf{U}^T(t,r) \cdot \mathbf{U}(t,r)\right] \cdot \Delta\mathbf{G}(r) \right\rangle dr \tag{B7}$$

and

$$R_A = \frac{1}{A_0^2} \int_{t_0}^{t} \left\langle \Delta\mathbf{G}(r) \cdot \left[\mathbf{U}^T(t,r) \cdot \mathbf{U}(t,r)\right] \cdot \mathbf{H}(r) \right\rangle dr. \tag{B8}$$

The analysis we now perform on $A_0^2$ will parallel the one we performed on $B_0^2$. Begin by expanding $\Delta\mathbf{G}$ in terms of the eigenvectors of $\left[\mathbf{U}^T(t,r) \cdot \mathbf{U}(t,r)\right]$

$$\Delta \mathbf{G} = \sum_{\beta=1}^{d} \Delta g^{(\beta)}(r,\tau)\ \hat{\mathbf{e}}^{(\beta)}(r,\tau).$$

Inserting this expression into Eq. (B7) results in

$$A_0^2 = \sum_{\beta=1}^{d} \int_0^{t-t_0} \left\langle \left[\Delta g^{(\beta)}(r,\tau)\right]^2 \exp\left[2\tau\lambda_\beta(r,\tau)\right]\right\rangle d\tau.$$

The integral shown above is exact. Before making assumptions/approximations similar to the ones listed above recall that the $v^{(\beta)}(r,\tau)$'s are completely determined by the eigenvectors, $\hat{\mathbf{e}}^{(\beta)}(r,\tau)$. In contrast the $\Delta g^{(\beta)}(r,\tau)$'s are associated with errors in modeling of the dynamics. Errors in the model of the dynamics are clearly independent of $\left[\mathbf{U}^T(t,r)\cdot\mathbf{U}(t,r)\right]$ and, thus, must be independent of its eigenvalues and eigenvectors. Therefore, the joint probability distribution of the $\Delta g^{(\beta)}(r,\tau)$'s and the $\lambda_\beta(r,\tau)$'s must factor. After factoring the joint probability distribution, utilizing assumptions/approximations similar to 2 and 3, and inserting Eq. (B4), we are left with

$$A_0^2 \simeq \left\langle|\Delta\mathbf{G}|^2\right\rangle \frac{1}{d}\sum_{\beta=1}^{d}\int_0^{\infty} \exp[2\tau\Lambda_\beta(\tau)]d\tau. \tag{B9}$$

The same type of crude analysis that resulted in Eq. (B6) produces the following expression for $R_A$

$$R_A \simeq \frac{\langle\Delta\mathbf{G}\cdot\mathbf{H}\rangle}{\langle|\Delta\mathbf{G}|^2\rangle} = \frac{1}{\langle|\Delta\mathbf{G}|^2\rangle}\left\langle\int_0^{r-} \Delta\mathbf{G}(r)\cdot\left[\mathbf{U}(r,\tau)\cdot\Delta\mathbf{G}(\tau)\right]d\tau\right\rangle \tag{B10}$$

This concludes our brief analysis of $B^2$ and $A^2$. Equations (B4), (B5), (B6), (B9), and (B10) represent the major theoretical expressions. Notice that all of these equations are functions of $\epsilon$, and, depending on how much the action of $\mathbf{U}(t,r)$ tends to rotate and shrink vectors, it is possible that $R_B$ and $R_A$ will not be positive. Finally, we remark that the integrals on the right hand side of Eq. (B9) are knowable in terms of the distributions of the local Lyapunov exponents of the model. Therefore, if $[1+2R_B]$ is of order one (which seems likely from Eq. (B10)) then by replacing $A_0$ in Eq. (B9) by an experimentally measured value of $A$ we can make an order of magnitude approximation for the modeling errors, $\left\langle|\Delta\mathbf{G}|^2\right\rangle$. It is important to note that this approximation can be made even though we do not know the true vector field for the dynamics.

In Figs. 13 we show $B$ vs $\epsilon$ for inband and Gaussian noise, and the models associated with DX2 and the vibrating wire. To obtain $B$ (as well as $A$) for a particular value of $\epsilon$ we produced curves like those shown in Figs. 7–9 and fitting these curves to Eq. (16) [54]. The figures indicate that when Gaussian noise is used $B$ is a much weaker function of $\epsilon$ than when inband noise is used. Since, $B = B_0$ for Guassian noise and $B = B_0[1+2R_B]^{1/2}$ for inband noise it follows that $B_0$ is a relatively weak function of $\epsilon$ when compared to $[1+2R_B]^{1/2}$. Indeed, since Figs. 13 indicates that $B_0$ is essentially constant under changes in $\epsilon$ we conclude that the functional dependence of $[1+2R_B]^{1/2}$ on $\epsilon$ is essentially the same as the functional dependence of $B$ on $\epsilon$. Quantitative details about the dependence of $R_B$ on $\epsilon$ are an intimate function of the type of noise that is found in the driving signal, and will change from one type of noise to the next. None the less, we expect $R_B$ to have a relatively strong dependence on $\epsilon$ when the autocorrelation function of the noise is not well approximated by a delta function. Finally, we notice that $B$ appears to become independent of $\epsilon$ when $\epsilon$ gets large, a result predicted by the analysis in Appendix A.

In Figs. 14 we show $A$ vs $\epsilon$ for the models associated with DX2 and the vibrating wire. We have determined $A$ for three distinct cases. The first and second cases, indicated by the circles and squares, used Gaussian and inband noise, respectively. The third case, indicated by diamonds, used the raw data as the driving term in Eq. (3) and the approximation $A^2 = \left\langle|\mathbf{z}|^2\right\rangle_T$. (For our experiments the noise levels are small so this is a valid approximation.) Recall that, although the noise types for raw data are unknown, they are probably not Guassian or inband. As one can see $A$ is not a function of the noise type, a result predicted by Eq. (15). By taking the ratio of Eqs. (B5) and (B9), and recalling that $B_0$ is essentially constanstant under changes in $\epsilon$ we conclude that the functional dependence of $A_0$ on $\epsilon$ is $A_0 = \mathcal{K}/\epsilon$, where $\mathcal{K}^2 = \left\langle|\Delta\mathbf{G}|^2\right\rangle/k(0)$ is a constant. A similar analysis indicates that the functional dependence of $R_A$ on $\epsilon$ is given by $\epsilon A/B = \mathcal{K}[1+2R_A]^{1/2}$ when Gaussian noise is present. In Figs. 15 we have plotted $\epsilon A/B$ (Gaussian noise) for the models associated with DX2 and the vibrating wire. We also notice that $A$ appears to become independent of $\epsilon$ when $\epsilon$ gets large, a result predicted by the analysis in Appendix A.

## APPENDIX C: THE DEPENDENCE OF $\Lambda_\beta(\tau)$ ON $\tau$

In this appendix we will examine the following approximation in detail

$$\int_0^{t-t_0} \langle \exp[2\tau\lambda_\beta(r,\tau)]\rangle \, d\tau \simeq \int_0^{\infty} \exp[2\tau\Lambda_\beta(\tau)] d\tau, \tag{C1}$$

where $\lambda_\beta(r,\tau)$ is the local Lyapunov exponent associated with evolution from $\mathbf{x}(r)$ to $\mathbf{x}(t)$. We will confirm that the integral on the right hand side is finite. The analysis will also confirm that the approximations associated with Eqs. (B1) and (B7) are accurate. If $2\tau\Lambda(\tau)$ falls off sufficiently fast for large $\tau$ then the upper limit of integration, $t - t_0$, can be replaced by $\infty$ and we will have verified that Eq. (C1) is a good approximation.

The remainder of this appendix is an analysis of the behavior of $\Lambda_\beta(\tau)$. It is tempting to begin an examination of $\Lambda_\beta(\tau)$ by modeling the distributions, $\rho_\beta(\lambda)$, by Gaussians. It is known that in the large $\tau$ limit the distributions, $\rho_\beta(\lambda)$, are well approximated by Gaussians [51] and

$$\langle \exp[2\tau\lambda_\beta(r,\tau)]\rangle = \exp\left[2\tau \langle \lambda_\beta(r,\tau)\rangle + 2\tau^2\sigma^2_{\lambda_\beta}\right],$$

where $\langle \lambda_\beta(r,\tau)\rangle$ is the mean value of $\lambda_\beta(r,\tau)$, and $\sigma^2_{\lambda_\beta}(\tau)$ is the variance of $\lambda_\beta(r,\tau)$ about the mean value [55].

In addition, it is known that in the $\tau \to \infty$ limit the mean values and variances of the local Lyapunov exponents converge as power laws, $\langle \lambda_\beta(r,\tau)\rangle \simeq \lambda_\beta + C_\beta/\tau$ (where $\lambda_\beta$ is the global Lyapunov exponent of Eq. (3) and $\sigma^2_{\lambda_\beta} \simeq D_\beta/\tau^{\nu_\beta}$ [23,51]. Thus, in this limit we can write

$$\Lambda_\beta(\tau) = \lambda_\beta + \frac{C_\beta}{\tau} + \frac{D_\beta}{\tau^{\nu_\beta - 1}}. \tag{C2}$$

(Despite this behavior it is important to remember that $\Lambda_\beta$ is *not* a Lyapunov exponent.)

However, in order to evaluated Eq. (C1) we also need to know the behavior of $\Lambda_\beta(\tau)$ for small $\tau$. In the small $\tau$ limit Eq. (C2) can not hold since $1/\tau$ would diverge. Furthermore, as stated in Appendix B, for small values of $\tau$ the distributions, $\rho_\beta(\lambda)$, are not Gaussians.

In Figs. 16 we have plotted some examples of $\Lambda_1(\tau)$, $\Lambda_2(\tau)$ and $\Lambda_3(\tau)$ as a function of $\tau$. A total of 10,000 initial conditions on the attractors were used for the ensemble averages. Figure 16a corresponds to results from the DX1 model with weak coupling ($\epsilon = 5$). When the coupling is increase the curves occur at lower values of $\Lambda(\tau)$ but retain their basic shape. Figure 16b corresponds to results from the vibrating wire model with strong coupling ($\epsilon = 3$). Similar results were also obtained for the DX2 model. A purely phenomenalogical model for the dependence of $\Lambda_\beta(\tau)$ on $\tau$ in the limit of small $\tau$ is $\Lambda_\beta(\tau) = A_\beta \exp(-\Gamma_\beta \tau^{\mu_\beta})$. By using a function that is zero for small $\tau$ and one for large $\tau$ we have constructed the following phenomenalogical function for $\Lambda_\beta(\tau)$

$$\Lambda_\beta(\tau) = A_\beta \exp[-\Gamma_\beta \tau^{\mu_\beta}] + [1 - \exp(-\Omega_\beta \tau)] \left[\lambda_\beta + \frac{C_\beta}{\tau} + \frac{D_\beta}{\tau^{\nu_\beta - 1}}\right]. \tag{C3}$$

The values of the global Lyapunov exponents, $\lambda_\beta$, are obtained by the usual methods for known equations of motion [51]. The parameters $A_\beta$, $\Gamma_\beta$, $\mu_\beta$, $\Omega_\beta$, $C_\beta$, $D_\beta$ and $\nu_\beta$ are obtained by using an annealing procedure to fit the data shown in Figs. 16 [54]. Other phenomenalogical function could be devised and we attach no significance to the functional form or the numerical values of the parameter other than that they produce a good fit to the data. The curves shown in Figs. 16a arise from graphing Eq. (C3) using the optimal parameter values. The dashed line indicates $\Lambda(\tau) = 0$ and is provided for the purposes of comparison.

In Figs. 17 we have plotted $\exp[2\tau\Lambda_\beta(\tau)]$ where the values of $\Lambda_\beta(\tau)$ are obtained from the data shown in Figs. 16. The curves shown in these figures are obtained by graphing $\exp[2\tau\Lambda_\beta(\tau)]$ using Eq. (C3). Figure 17a indicates that the integrands in Eq. (C1) fall off quickly to zero which implies that the integrals don't diverge. When strong coupling is used for this model (the DX1 model) the curves drop off even faster with increasing values of $\tau$. In Fig. 17b we show the corresponding results for the vibrating wire model. In this figure $\exp[2\tau\Lambda_1(\tau)]$ takes longer than $\tau = 4$ to decrease to a value near zero, however an inspection of Eq. (C3) indicates that it will eventually approach zero. The fact that none of the integrals in Eq. (C1) diverges can be analytically verified by inserting Eq. (C3) into Eq. (C1) and considering the large $\tau$ limit of the integrand.

---

FIG. 1. The distances between the orbits **x**, **y**, and **w**. The dashed line is $\mathbf{z}^2 = |\mathbf{x} - \mathbf{y}|^2$ while the solid line is $\mathbf{z}^2 = |\mathbf{x} - \mathbf{w}|^2$. Near synchronization between **x** and **y** is clearly demonstrated. The driving variable is $x_3 + \sigma \hat{u}_3$, $\epsilon = 20$, and the data is from $\alpha = 17.4$. The time step is in units of $\Delta t = 0.02$

FIG. 2. A schematic of a piece of the trajectory, **x**. The true vector field, **G**, the model vector field, **F**, and the difference $\Delta\mathbf{G}$ are shown.

FIG. 3. The block diagram of the circuit we have used. For our numerical tests we vary the parameter $\alpha$.

FIG. 4. The attractors corresponding to data sets obtained from the electronic circuit. The data vectors are formed by using a time delay embedding of the scalar signals. (a) DX1. (b) DX2.

FIG. 5. The attractors corresponding to the data set obtained from the vibrating wire. The data vectors are formed by using a time delay embedding of the scalar signal.

FIG. 6. The attractors corresponding to the DX2 data set with additive noise of magnitude $\sigma = 1$. (a) Gaussian noise. (b) Inband noise.

FIG. 7. The normalized synchronization deviation levels for the DX1 model as a function of added noise.

FIG. 8. The normalized synchronization deviation levels for the DX2 model as a function of added noise.

FIG. 9. The normalized synchronization deviation levels for the vibrating wire model as a function of added noise.

FIG. 10. An example of a bust that occurred when synchronizing DX1 to its model. The driving signal has been contaminated with inband noise of size $\sigma = 1$, and coupling strength $\epsilon = 5$.

FIG. 11. The synchronization level as a function of $\Delta\alpha$ and the coupling strength, $\epsilon$. When $\Delta\alpha = 0$ the driving signal has the same dynamics as the data used to construct the model. Each change in $\Delta\alpha$ represents an change of approximately 1 % in the value of $\alpha$.

FIG. 12. The densities of the local Lyapunov exponents for the DX1 model with $\epsilon = 5$ and $\tau = \Delta t = 0.02$. There are three densities. The one on the right is associated with $\lambda_1$, the one in the middle is associated with $\lambda_2$, and the one on the left is associated with $\lambda_3$.

FIG. 13. A graph of $B$ vs $\epsilon$ for Gaussian and inband noise for models associated with DX1 and the vibrating wire. The circles and squares are associated with Gaussian and inband noise, respectively. (a) The DX2 model. (b) The vibrating wire model.

FIG. 14. A graph of $A$ vs $\epsilon$. We have calculated $A$ using raw data (the diamonds) data with Gaussian noise (the circles) and data with inband noise (the squares). The figure indicates that $A$ is independent of the noise type. (a) The DX2 model. (b) The vibrating wire model.

FIG. 15. A graph of the ratio $\epsilon A/B$ for Gaussian noise. This quantity is proportional to $[1+2R_A]^{1/2}$. The circles and squares correspond to DX1 and DX2 respectively. Notice that for $\epsilon \geq 5$ the results for DX1 and DX2 indicate that $R_A$ is essentially a quadratic function of $\epsilon$.

FIG. 16. The values of $\Lambda_\beta(\tau)$ as a function of the evolution time, $\tau$. The $\Lambda_\beta(\tau)$'s approach the global Lyapunov exponents as power laws for large $\tau$. The circles, squares, and triangles correspond to $\Lambda_1$, $\Lambda_2$, and $\Lambda_3$, respectively. (a) The DX1 model for weak coupling ($\epsilon = 5$). (b) The vibrating wire model for strong coupling ($\epsilon = 3$).

FIG. 17. The values of $\exp[2\tau\Lambda_\beta(\tau)]$ as a function of the evolution time, $\tau$. The circles, squares, and triangles correspond to $\Lambda_1$, $\Lambda_2$, and $\Lambda_3$, respectively. (a) The results for the DX1 model and $\epsilon = 5$. (b) The results for the vibrating wire model and $\epsilon = 3$.